\documentclass[12pt]{article}

\usepackage{scicite}
\usepackage{graphicx}
\usepackage{subfigure}
\usepackage{multirow}
\usepackage{amssymb}
\usepackage{soul}
\usepackage{caption}
\usepackage{tabulary,amsmath}
\usepackage{times}

\newenvironment{sciabstract}{%
\begin{quote} \bf}
{\end{quote}}

\newcounter{lastnote}

\title{Knowledge Discovery in Nanophotonics Using Geometric Deep Learning}

\usepackage{authblk}

\author[1]{Yashar Kiarashinejad}
\author[1]{Mohammadreza Zandehshahvar}
\author[1]{Sajjad Abdollahramezani}
\author[1]{Omid Hemmatyar}
\author[ ]{Reza Pourabolghasem}
\author[1]{Ali Adibi\thanks{Corresponding author: ali.adibi@ece.gatech.edu}}
\affil[1]{Georgia Institute of Technology, 778 Atlantic Drive NW, Atlanta, 30332, GA, USA}
\date{}

\begin{document} 


\baselineskip24pt


\maketitle


\begin{sciabstract}
We present here a distinctive approach for using the intelligence aspects of artificial intelligence for knowledge discovery rather than the conventional task of device optimization in electromagnetic (EM) nanostructures. This approach uses training data obtained through full-wave EM simulations of a series of nanostructures to train geometric deep learning algorithms to assess the range of feasible responses as well as the feasibility of a desired response from a class of nanophotonic structures. To facilitate the knowledge discovery and reduce the computation complexity, our approach combines the dimensionality reduction technique (using an autoencoder) with convex-hull and one-class support-vector-machine (SVM) algorithms to find the range of the feasible responses in the latent (or the reduced) response space of the EM nanostructure. We show that by using a small set of training instances (compared to all possible structures), our approach can provide better than 95\% accuracy in assessing the feasibility of a given response. More importantly, the one-class SVM algorithm can be trained to provide the degree of feasibility (or unfeasibility) of a response from a given nanostructure. This important information can be used to modify the initial structure to an alternative one that can enable an initially unfeasible response. To show the applicability of our approach, we apply it to two important classes of binary metasurfaces (MSs), formed by an array of plasmonic nanostructures, and periodic MSs formed by an array of dielectric nanopillars. In addition to theoretical results, we show the experimental results obtained by fabricating several MSs of the second class. Our theoretical and experimental results confirm the unique features of this approach for knowledge discovery in nanophotonics applications.
\end{sciabstract}

% In setting up this template for *Science* papers, we've used both
% the \section* command and the \paragraph* command for topical
% divisions.  Which you use will of course depend on the type of paper
% you're writing.  Review Articles tend to have displayed headings, for
% which \section* is more appropriate; Research Articles, when they have
% formal topical divisions at all, tend to signal them with bold text
% that runs into the paragraph, for which \paragraph* is the right
% choice.  Either way, use the asterisk (*) modifier, as shown, to

\section{Introduction}

Photonic nanostructures have been of great recent interest due to their unique capabilities to manipulate the properties of electromagnetic (EM) waves beyond what conventional bulk materials can do. Owing to their constituent nanoscale features, which can spectrally, spatially, or temporally control the optical state of EM waves with subwavelength resolution, nanophotonic devices offer all the functionalities realized by conventional bulky optical devices in much smaller footprints \cite{ding2017gradient,kamali2018review,ding2019dynamic,genevet2017recent,jahani2016all,zhan2016low,jiang2019metasurface,abdollahramezani2018dynamic,hemmatyar2017phase,abdollahramezani2018reconfigurable,abdollahramezani2015beam,abdollahramezani2015analog}. Combined with the advances in nanofabrication technologies, these nanostructures have been used to demonstrate devices with enormous potential for groundbreaking technologies such as computing \cite{chizari2016analog,shen2017deep,zhu2017plasmonic,abdollahramezani2017dielectric}, imaging \cite{colburn2018metasurface}, and energy harvesting \cite{liu2011taming,ding2014ultrabroadband},as well as electronics\cite{rashidi2018improving,rashidi2019survey}. 

%% to name a few

\begin{figure}[htbp]
	\centering
	\includegraphics[trim=0cm 0cm 0cm 0cm,width=1\textwidth,clip]{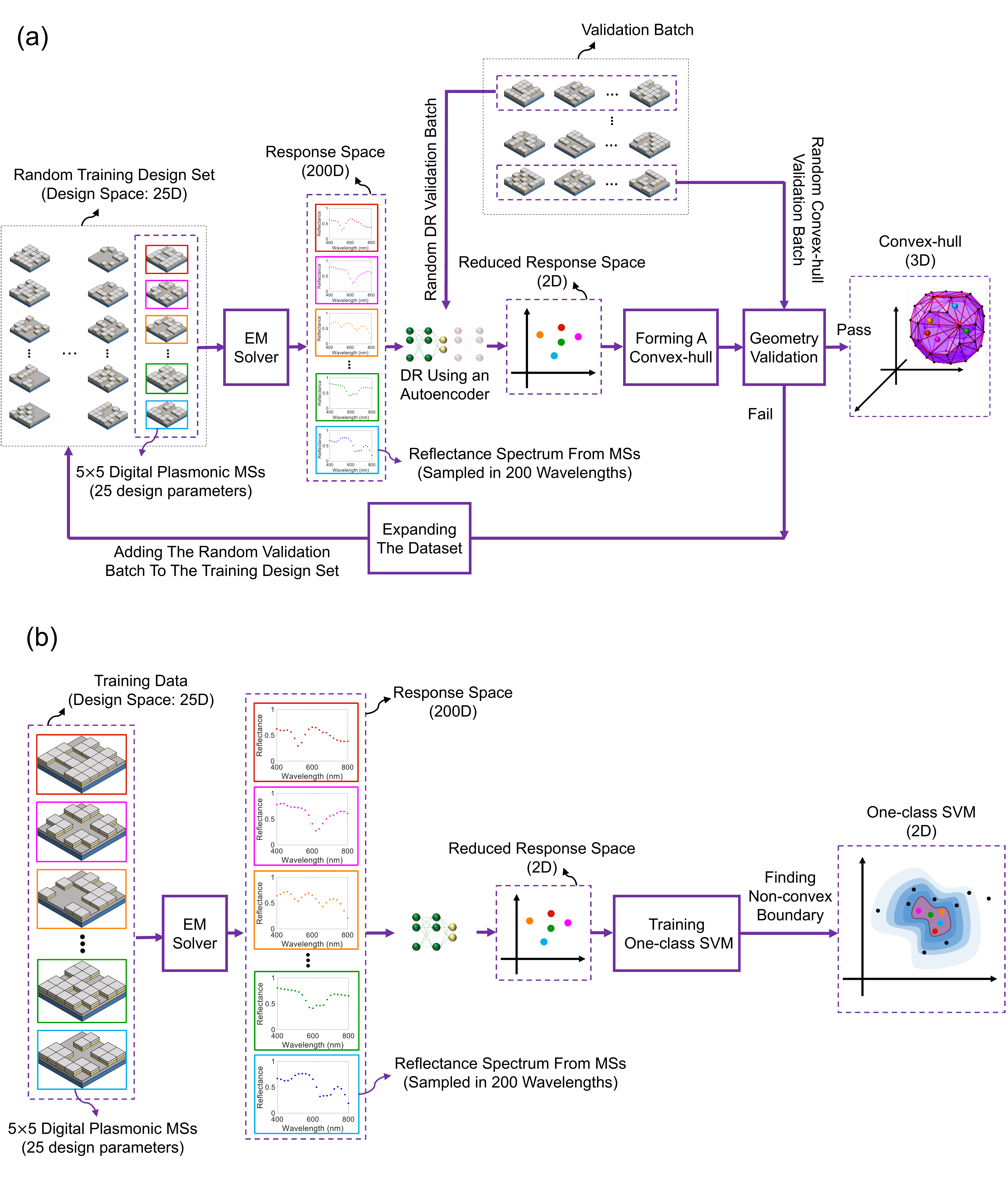}
	\caption{(a) Training algorithm for finding the convex-hull of the patterns in the latent RS. (b) Using one-class SVM over patterns in the latent RS to investigate the level of feasibility of a desired response. The dimensionality of the latent RS is found by training the autoencoder. The 2D and 3D representations are just examples for facilitating graphical understanding.}
	\label{fig:block diargram presented technique}
\end{figure}

Design of photonic devices in the nanoscale regime outperforming the bulky optical components has been a long-lasting challenge in state-of-the-art applications. Accordingly, devising a comprehensive model to understand and explain the fundamental physics of light-matter interaction in these nanostructures is a substantial step toward the realization of novel nanophotonic devices. To this end, existing modeling methods can be categorized into two main groups; single- and multi-objective approaches. Single-objective approaches either rely on exhaustive design parameter sweeps using a brute-force EM solver (e.g., based on the finite element method (FEM)) \cite{wu2003design} or evolve from an initial guess to a final result through evolutionary methods (e.g., genetic algorithm) \cite{bossard2014near}. While the former requires extensive computation, the latter highly depends on the initial guess and in most cases converges to a local optimum. Both of these single-objective approaches are computationally demanding and fail when the input-output relation is complex, or the number of desired features for a nanostructure grows. On the other hand, multi-objective methods \cite{jiang2019global,kiarashinejad2019deepdim} deal with formation of a model to optimize a certain class of problems. Although these methods are more computationally efficient, obtaining an optimal solution is not guaranteed.

Deep learning (DL)-based design approaches, combined with limited exhaustive searches, have proven to be a potent solver of multi-objective optimization problems by learning the input-output relation \cite{jiang2019dataless,jiang2019free,yao2019intelligent,tahersima2019deep,ma2 018deep,campbell2019review,sajedian2019optimisation,baxter2019plasmonic,long2019inverse,an2019generative,kudyshev2019rapid,liu2018generative,qiu2019deep}. DL-based approaches combined with dimensionality reduction algorithms have been recently developed for design and optimization of EM nanostructures\cite{kiarashinejad2019deepdim,kiarashinejad2019deep,hemmatyar2019full,kiarashinejad2019mitigating,KiarashinejadFIO2019,HemmatyarFIO2019,ZandehshahvarFIO2019}. More importantly, such novel techniques can provide considerably valuable insight about the dynamics of light-matter interaction in nanostructures with the hope of uncovering new physical phenomena that can be used to form completely new types of devices. Despite initial proof of principle\cite{kiarashinejad2019deep}, there has been little effort on rigorously and systematically using these techniques to obtain detailed knowledge about the physics of light-matter interaction in EM nanostructures (e.g., metasurfaces (MSs)). The change of focus of using DL techniques from “optimization” to “knowledge discovery” can open a new research area with potentially transformative results in the entire field of nanophotonics. Examples of these “knowledge discovery” paradigms include assessing the feasibility of a desired response using a given structure as well as the range of possible responses a given design can provide. Since existing optimization and inverse design approaches provide a solution to any inverse design problem and, to the best of our knowledge, such approaches have not considered the important concept of design feasibility, we believe our proposed method can pave the way for more efficient and practical and fabricationally favorable design paradigms. 
Knowing the feasibility of a desired response offered by a photonic nanostructure is very helpful prior to any design or optimization effort in avoiding suboptimal designs or convergence issues. It also guides us to modify the initial structure to achieve the desired response. To the best of our knowledge, this important concept has not been considered in existing optimization and inverse design approaches, which provide a solution to any inverse design problem regardless of its feasibility.  

In this paper, we present a new geometric deep learning (GDL)-based technique by forming the smallest convex set (i.e., the convex-hull)\cite{boyd2004convex} to discover hidden optical phenomena while analyzing the feasibility of having a desired optical response from a certain class of EM nanostructures. GDL is a term used for techniques that aim to generalize DL approaches by considering the non-Euclidean domain such as manifolds. These methods reduce the dimensionality of the discovered patterns in the design and response space while ﬁnding the governing geometry of such patterns in lower-dimensional space (reduced space) in which the Euclidean distance can be a good measure for similarity of different patterns \cite{meilua2007comparing,chen2019selecting,perraul2013non,patrikainen2006comparing, jin2018representing}. The developed approach in this paper is based on reducing the dimensionality of the response space (RS) of a given EM nanostructure and finding the convex-hull that contains achievable responses in the reduced RS (also known as the latent RS). The dimensionality reduction (DR) implementation is based on an autoencoder \cite{hinton2006reducing}, and the Quickhull \cite{barber1996quickhull} algorithm is used to form the convex-hull in the latent RS. Our technique uses the numerical simulation of the response of the system for a series of randomly selected design parameters (called training set) and another series of similar simulations for validation of the technique. After initial training and validation, the algorithm finds the optimal bounded subset, which contains all feasible responses.

The optimal region that contains the feasible responses might not be convex in many cases, and it is better to also find a tighter bound over feasible responses in the latent RS. For this purpose, we use the one-class support vector machine (SVM) algorithm \cite{scholkopf2000support} to find the non-convex geometry. One-class SVM also provides information about the level of feasibility (or unfeasibility) of a response and the possibility of trading an acceptable error (or a small change in the desired response) to get the closest feasible response to a unfeasible one (desired). Despite being implemented for the EM nanostructures (especially dielectric and plasmonic MSs), our technique can be applied to a wide variety of applications once the training data can be provided. Some example extensions include thermal structures, fluidic systems, mechanical platforms, and acoustic metamaterials. 

The rest of the paper is organized as follows. Section 2 describes the details of the GDL-based approach. Section 3 demonstrates the application of the approach to two classes of important MSs. Section 4 is devoted to the comparison of the findings of our technique with experimental data. It is followed by the further discussions in Section 5 and conclusion in Section 6.

\section{Theoretical Framework}
\subsection{Convex-hull of a Set of Data Points}
The convex-hull of a set of points is defined as the smallest convex set that contains all those points\cite{boyd2004convex}. A $d$-dimensional convex-hull can be represented using its vertices and ($d-1$)-dimensional facets. The ridges of the convex-hull are ($d-2$)-faces, which are the intersections of the vertices in two neighboring facets. There are different algorithms presented in geometrical computation to form the convex-hull of a given set of points. One of the most effective and well-known algorithms is Quickhull, which forms the convex-hull using an incremental method based on Grunbaum's Beneath-Beyond theorem (see the Supplementary Information Section 1)\cite{barber1996quickhull}. For a typical problem, the Quickhull algorithm starts with a set of given (or training) points and forms the initial convex-hull. The points that lie outside the initial convex-hull are considered as the outside set. The farthest point from the initial convex-hull (i.e., the point with the maximum Euclidean distance from its nearest facet) is found at each iteration, and the facets, ridges, and vertices are updated based on Grunbaum's Beneath-Beyond theorem\cite{barber1996quickhull}. These steps are repeated until the algorithm converges. 

While the convex-hull algorithm is capable of finding a convex geometry for feasible responses, it has some limitations. If the optimum feasible region is not convex, inevitably some unfeasible regions in the latent RS will be included in the convex-hull to reach a convex region. This limits the efficiency of the algorithm for such structures due to the false-positive errors. Moreover, the algorithm acts as a binary classifier and classifies responses into two classes: feasible (achievable) and unfeasible (unachievable). In most practical cases, it is desirable to know how far an unfeasible response is from feasible responses. It is also helpful to know whether it is possible to push an unfeasible response toward the feasible region by accepting some error. Unfortunately, the Euclidean distance of a given point in the latent RS from the boundaries of the convex-hull is not a good measure for feasibility of the corresponding response. To address this limitation, we use one-class SVM in the latent RS as the alternative algorithm. 

\subsection{One-class SVM for a Set of Data Points}

One-class SVM is an algorithm that separates the patterns into two regions (e.g., feasible and unfeasible in our case). In addition, the Euclidean distance between any point in the space and the boundaries of the one-class SVM is a good measure of this separation (e.g., a good measure of the feasibility of a response in our case). Mathematically, a one-class SVM forms a nonlinear geometry by projecting patterns $x_i$ through a nonlinear function $\phi$ to a higher-dimensional space $F$. This mapping helps to separate linearly non-separable patterns in the low-dimensional input space $I$ in  high-dimensional space by $F$a hyperplane (represented with $w^T + b = 0$, $w \in F$ and $b \in \mathbb{R}$). By projecting this hyperplane from the high-dimensional space back to the original space, the algorithm finds the equivalent non-convex decision geometry. In this projection, the resulting region for the desired (or feasible) class of data may not only have a non-convex geometry, but it may also exclude smaller closed regions within the geometry. The implementation of the one-class SVM has considerable flexibility through two parameters $\nu$ and $\gamma$ which control the tightness of the geometry of the decision region and the maximum ratio of the given training patterns that fall outside the geometry (and thus, contribute to the classification error). By using different values of $\gamma$, one can find a series of boundaries with different levels of classification errors for the ground-truth data. Although one-class SVM is capable of finding the non-convex geometry of latent patterns, computation complexity of validating $\nu$ and $\gamma$ in each iteration prevents using it as a prelimenary approach of forming the geometry in many cases. Using the convex-hull forming algorithm to get the initial information for implementing the one-class SVM algorithm is a very effective solution to this challenge. Further details about the one-class SVM are provided in the Supplementary Information section S2.

\subsection{Investigation of the Feasibility of a Desired Response from a Given Structure}

\begin{figure}
\centering
\includegraphics[width=1\linewidth, trim={0cm 0cm 0cm 0cm},clip]{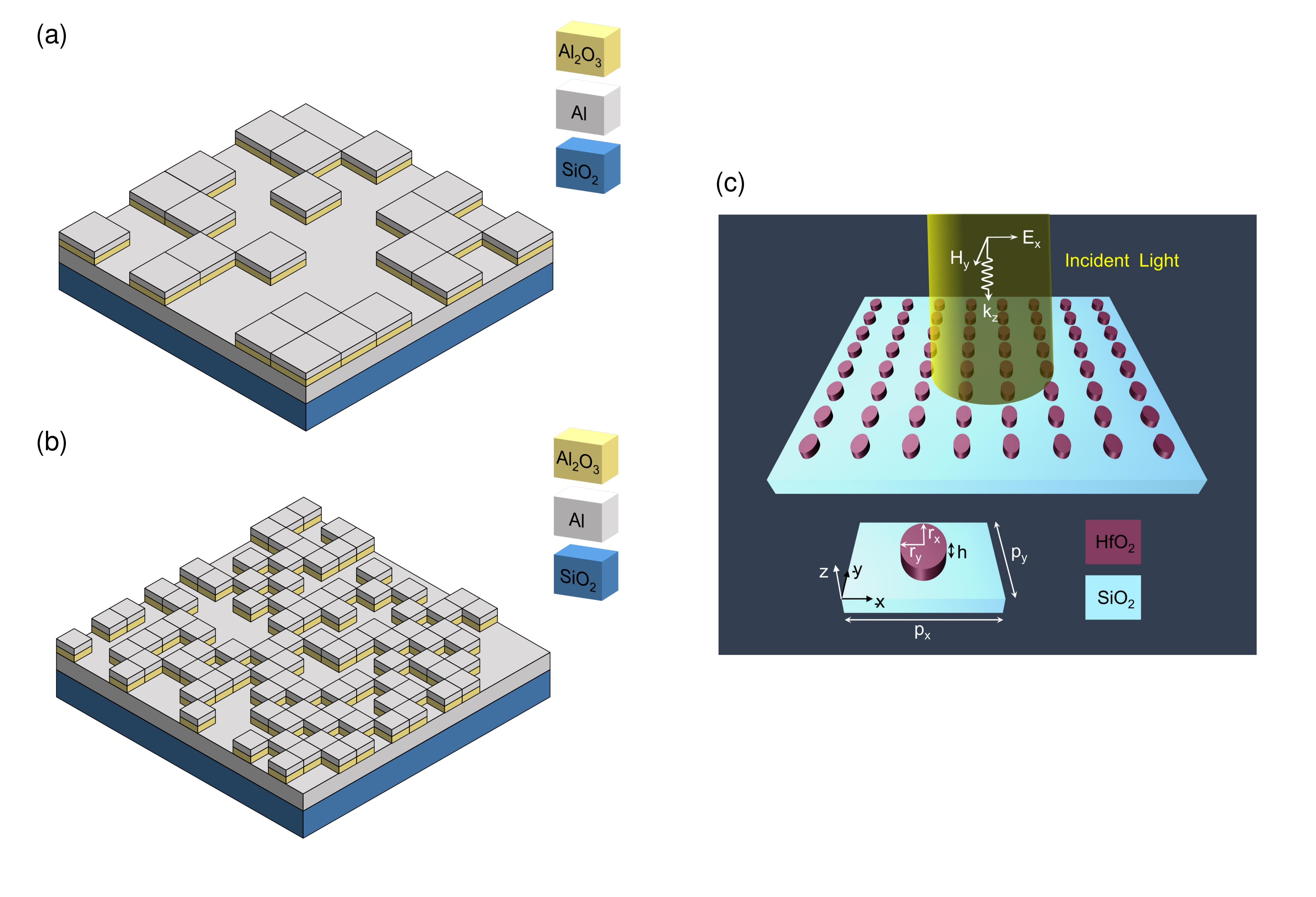}
\caption{(a), (b) The schematic of a unit cell of a digital plasmonic MS composed of a square lattice of (a) 7$\times$7, and (b) 14$\times$14 binary inclusions. The substrate consists of an Al layer as the back-reflector covered by an array of Al/Al$_2$O$_3$ binary inclusions. The thickness of the deposited Al back-reflector, Al, and Al$_2$O$_3$ layers in both cases are 100 nm, 35 nm, and 35 nm, respectively. (c) The unit cell of a median-index dielectric MS consisting of HfO$_2$ nanopillars, deposited using atomic layer deposition (ALD) with the optical properties reported in \cite{devlin2016broadband}. The structure is normally illuminated by a TM- polarized EM wave (TM: transverse magnetic, i.e., magnetic field in the plane of the inclusions), as shown in (c), and the reflection response is calculated and sampled over the bandwidth of 400-800 nm to be introduced to the algorithms in Fig.~\ref{fig:block diargram presented technique}.}
\label{fig:nanostructures}
\end{figure}

Figure~\ref{fig:block diargram presented technique}(a) shows the schematic of our technique for forming the convex-hull for the feasible responses of a given nanostructure. In the first step, a full-wave EM simulation software (or alternatively an EM wave solver using an analytic or a semi-analytic model) is used to provide an initial batch of randomly generated patterns (we refer to them as the input dataset). Each pattern is calculated using a given set of randomly selected design parameters (i.e., a point in the design space (DS)), and thus, it relates the DS to the RS. Then, we reduce the dimensionality (see Supplementary Information, Section S3) of the RS by training an autoencoder using a subset of the available training data and a desired autoencoder reconstruction error. Next, we use the Quickhull \cite{barber1996quickhull} algorithm to form a convex-hull to bound the patterns in the latent RS. Then, we validate the convex-hull by using a batch of validation data. Since all of the validation responses originate from a feasible structure, the optimum convex-hull should bound all the validation data. We put a threshold for the validation success rate. If the convex-hull does not pass the validation step, the validation batch will be added to the initial training batch to expand the training dataset for retraining the algorithm. This process is repeated until the resulting convex geometry reaches the desired validation success rate. After convergence, the convex geometry is tested using an unseen test dataset (that includes both feasible and unfeasible responses) to find its performance defined by the error rate. A similar process is used for training the one-class SVM as shown in Fig.~\ref{fig:block diargram presented technique}(b) to find the non-convex geometry of feasible response patterns in the latent RS.

\section{Response Feasibility Investigation}

To demonstrate the potentials of our technique, we apply it to the investigation of possible optical reflection responses from plasmonic and dielectric MSs as two popular classes of photonic nanostructures. Figures~\ref{fig:nanostructures}(a) and \ref{fig:nanostructures}(b) show two implementations of a digital plasmonic MS consisting a 7$\times$7 and a 14$\times$14 array of binary nanocubes of stacked aluminum/alumina (Al/Al$_2$O$_3$), respectively. The significant number of plasmonic inclusions or design parameters, especially that in Fig.~\ref{fig:nanostructures}(b), allows these structures  to form sophisticated EM responses like Fano and Lorentzian resonances \cite{fano1961effects,limonov2017fano}. As an alternative, we also consider a median-index dielectric MS formed by a square lattice array of hafnia (HfO$_2$) nanoparticles on a transparent substrate as shown in Fig.~\ref{fig:nanostructures}(c). For both classes of MSs, we train a convex-hull and a one-class SVM to quantitatively evaluate the practical feasibility of any desired response based on a small set of simulation results found by calculating the reflection spectrum of the MS in the visible wavelength range using the FEM implemented in a commercial full-wave package COMSOL Multiphysics as explained in the Methods.

% Table~\ref{tab:distance_6-D} shows example feasible responses from the structures in 
\begin{figure}
\centering
\includegraphics[width=0.9\linewidth, trim={1cm 3cm 2cm 2cm},clip]{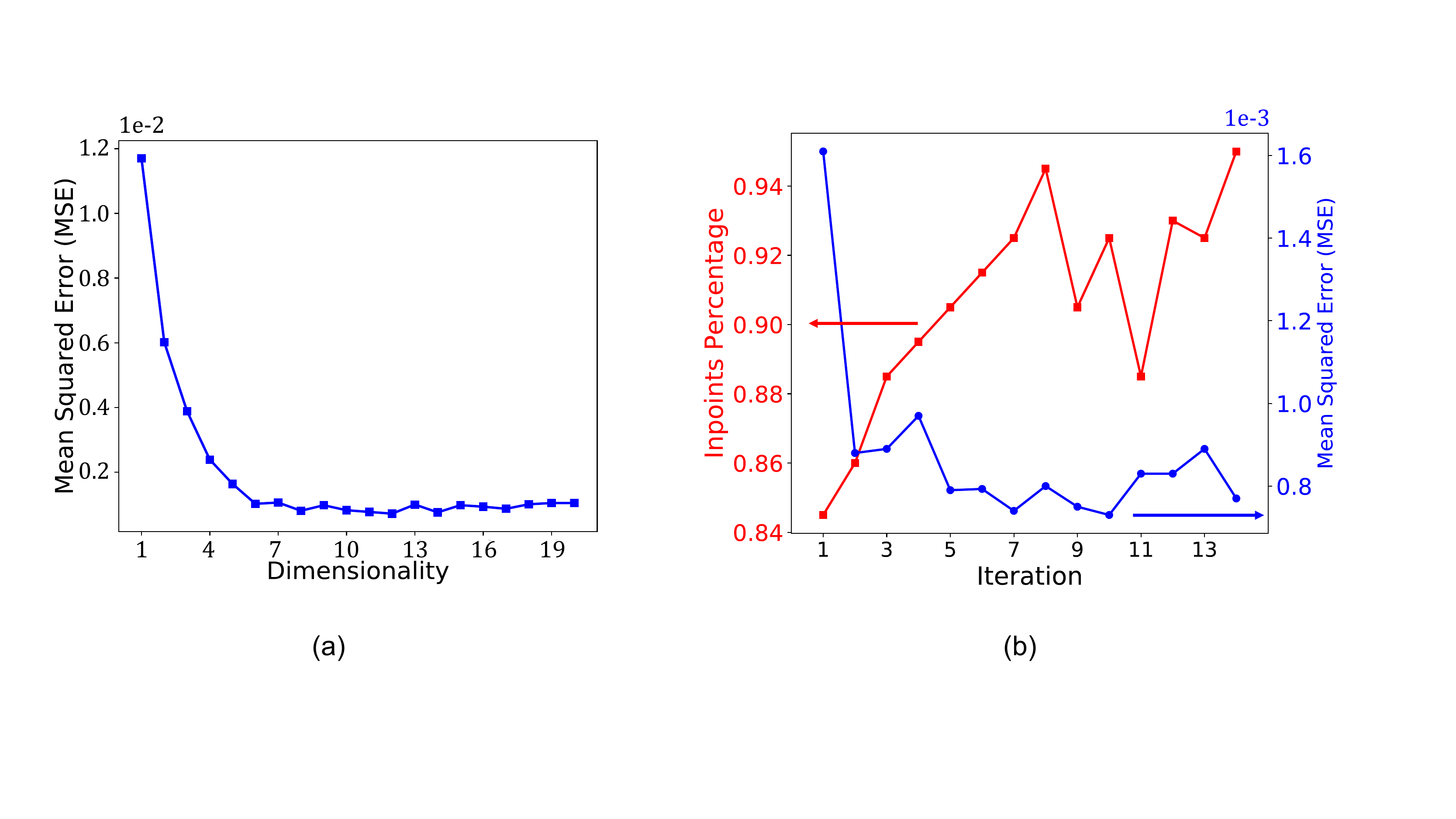}

% \subfigure[]{\includegraphics[width=0.49\linewidth, trim={0cm 0cm 0cm 0cm},clip]{Inpoints_Train_6d.pdf}}
% \subfigure[]{\includegraphics[width=0.45\linewidth, trim={0cm 0cm 0cm 0cm},clip]{MSE_Train_3d.pdf}}
\caption{(a) Reconstruction MSE for autoencoder trained on the responses of the 14$\times$14 binary structure in Fig. \ref{fig:nanostructures}(a)  for different dimensionalities of the latent RS. Using the results, we select 6 as the desired dimensionality of the latent RS. Responses can be reconstructed after reducing dimensionality from 200 to 6 by accepting less than 5$\%$ error. Examples of reconstructed responses and the ground truth responses are presented in the Supplementary Information Section 4. (b) Auto-encoder training error and in-points percentage for the algorithm in Fig. \ref{fig:block diargram presented technique}(a) for 14$\times$14 binary structure in Fig. \ref{fig:nanostructures}(a) after different iterations of the algorithm. The algorithm converged after 14 iterations.} 
\label{fig:MSE_convexhull_results}
\end{figure}
% Figures~\ref{fig:nanostructures}(a) and \ref{fig:nanostructures}(b) show two examples of binary structures (14$\times$14 and 7$\times$7) . 
The design patterns in each case are achieved by random selection of the binary inclusions, and the calculated reflection spectra are sampled uniformly in the 400-800 nm wavelength range with 2 nm resolution to form a vector with dimensionality of 200 as the response pattern. Due to the iterative nature of the algorithm in Fig.~\ref{fig:block diargram presented technique}(a), the minimum number of training data depends on the number of iterations for convergence. In addition, we use 500 simulated response patterns for testing the algorithms after convergence. Based on several simulations to understand the requirement of the selected structure, we chose 8000 as the size of the training/validation dataset. Knowing that achieving an ideal Fano lineshape is not possible with these structures (due to remarkable Ohmic loss of metals in the visible range), we also formed 80 ideal Fano lineshapes over the 400-800 nm spectral using the Equation S8 (see Supplementary Information section 5) as unfeasible responses to test the algorithms. 

% Figure \ref{fig:convexhull_results}(c) shows a few examples of these unfeasible responses. 

After obtaining the training dataset, the first step of implementation is the DR of the RS by training an autoencoder. To find the optimum dimensionality of the latent RS and the number of layers of the autoencoder, we use an ad-hoc approach by using different autoencoder structures and dimensionalities and calculating the mean squared error (MSE) for each case. The details of this approach is explained in Ref. \cite{kiarashinejad2019deep} and not repeated here. Figure~\ref{fig:MSE_convexhull_results}(a) shows the variation of the MSE of the autoencoder trained for the 14$\times$14 array in Fig.~\ref{fig:nanostructures}(b) with the dimensionality of the latent RS. The autoencoder in each case is compared of 7 layers with 200,100,50,X,50, 100,and 200 neurons with X being the dimensionality of the latent RS. Training and testing the autoencoder is performed with 8000 and 2000 random response patterns, respectively. Figure~\ref{fig:MSE_convexhull_results}(a) suggests that using 6 as the dimensionality of the latent RS results in MSE of 0.001, which can be translated to less than 5$\%$ point-to-point error (see Supplementary Information, Section S3). 

It is important to note that the goal of this initial training step is to find the optimum dimensionality of the autoencoder in Fig.~\ref{fig:block diargram presented technique}. For optimal training of either algorithms in Figs.~\ref{fig:block diargram presented technique}(a) and (b), we use an untrained autoencoder with the optimum dimensionality and train the entire algorithm (composed of the autoencoder follows by the Quickhull to form the convex geometry). To find the optimum convex-hull in the resulting latent RS, we start with an initial batch of data with 5000 ground-truth patterns in the algorithm in Fig.~\ref{fig:block diargram presented technique}(a) (with dimensionality of RS being 6) to train the cascaded autoencoder and forming the convex-hull in the 6-dimensional latent RS space. At each iteration, we use 200 validation data (without replacement) for autoencoder and 200 for convex-hull. We select 0.001 (5$\%$ point-to-point error) for the autoencoder validation threshold and 95$\%$ for in-point percentage (i.e. percentage of the ground truth patterns lies inside the boundary), respectively. The algorithm converged after 14 iterations. As a result, we used 11000 data to reach convergence.

\begin{figure}
\centering
% \subfigure[]{\includegraphics[width=0.45\linewidth, trim={0cm 0cm 0cm 0cm},clip]{ConvHull_Train_2d.pdf}}
\subfigure[]{\includegraphics[width=0.49\linewidth, trim={0cm 0cm 0cm 0cm},clip]{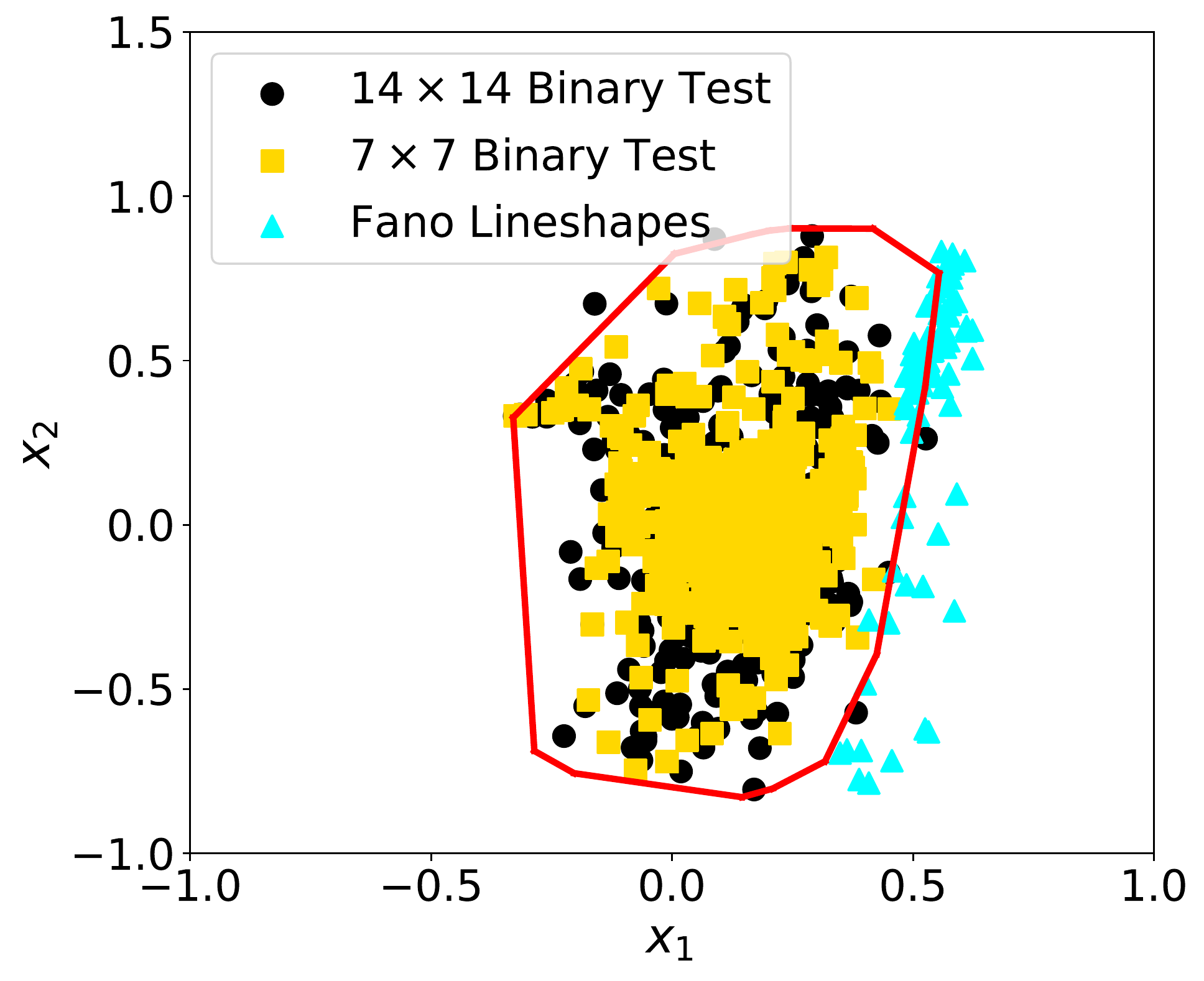}}
\subfigure[]{\includegraphics[width=0.49\linewidth, trim={0cm 0cm 0cm 0cm},clip]{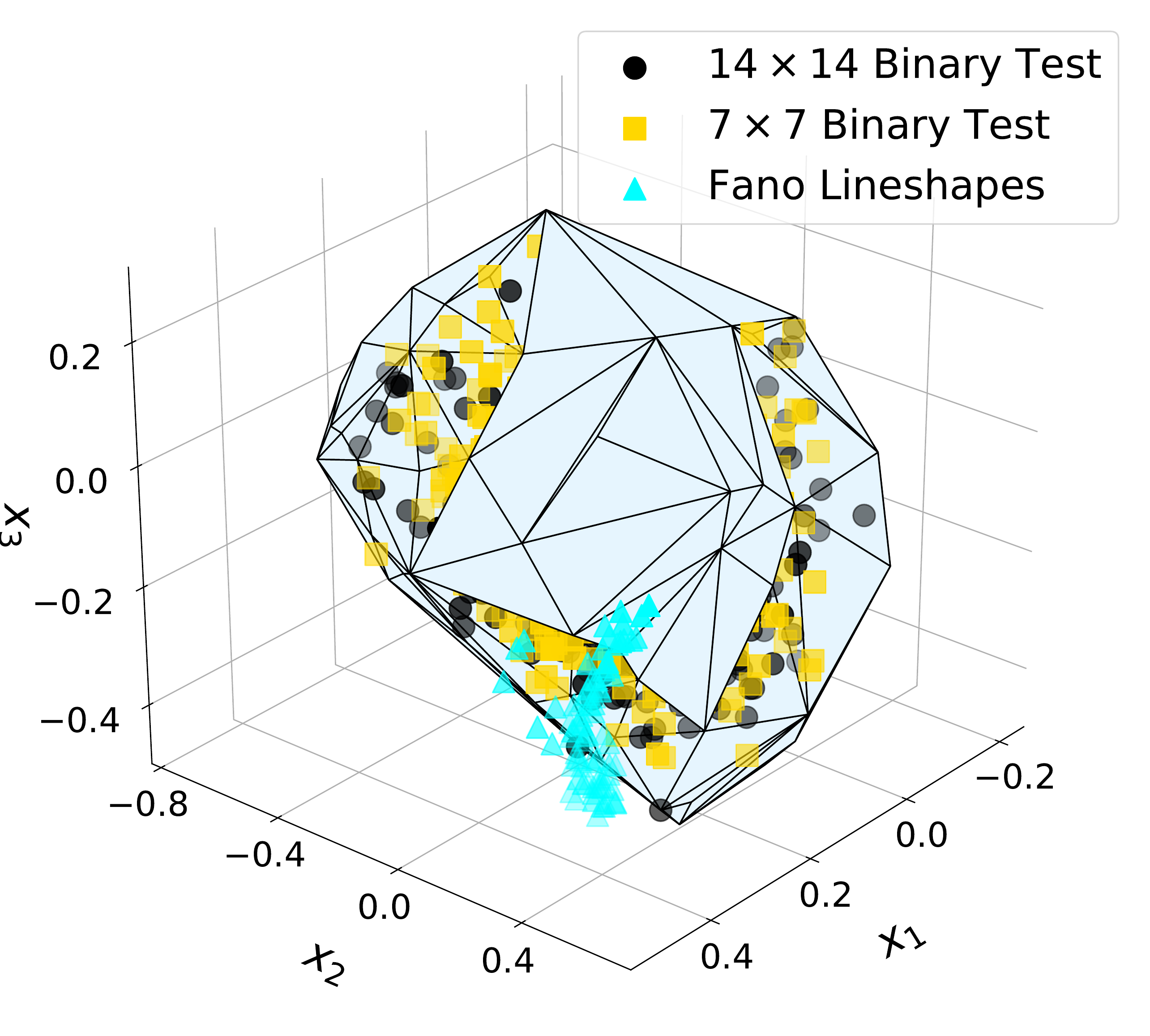}}
\subfigure[]{\includegraphics[width=0.49\linewidth, trim={0cm 0cm 0cm 0cm},clip]{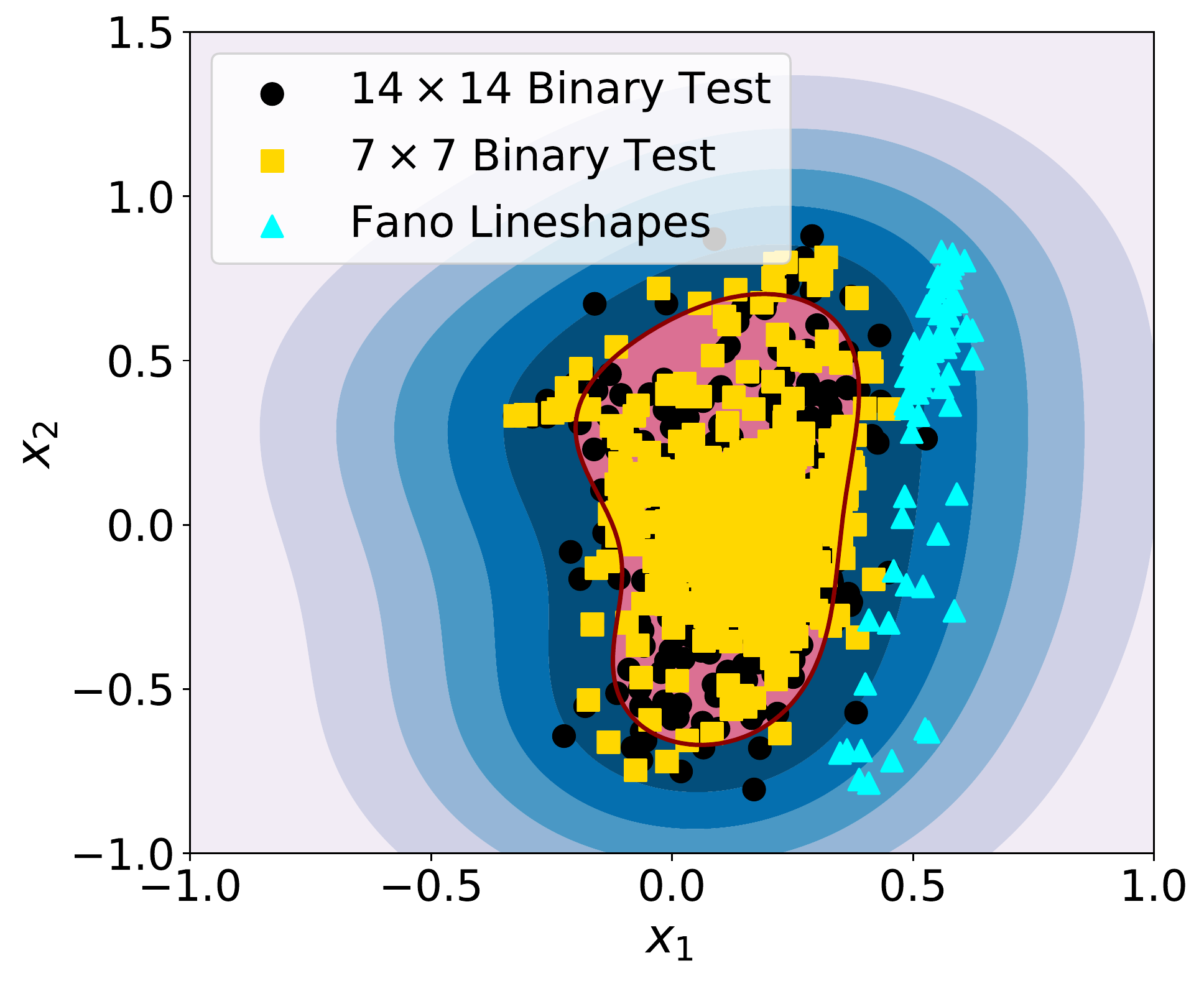}}
\caption{Representation of the convex-hulls in (a) 2D and (b) 3D RS space of the 14$\times$14 binary structure in Fig.~\ref{fig:nanostructures}(b). The feasible responses for the 14$\times$14 and 7$\times$7 binary structures and the unfeasible ideal Fano lineshapes are shown. (c) Non-convex geometry for the feasible responses found by one-class SVM algorithm along with feasible and unfeasible responses in the 2D latent RS for the 14$\times$14 binary structure in Fig.~\ref{fig:nanostructures}(b).}
\label{fig:convexhull_results}
\end{figure}

Figure~\ref{fig:MSE_convexhull_results}(b) shows the MSE of the autoencoder and percentage of ground truth test data that lie in the convex-hull after each iteration. After validating the convex-hull and its corresponding autoencoder, we feed our test data consisting of feasible responses for the 14$\times$14 and 7$\times$7 binary nanostructures as well as unfeasible ideal Fano resonances to the algorithm. The results (see Table~\ref{tab:inpoint_percentage}) show that about 91.8$\%$ of the feasible responses of the 14$\times$14 structures, 96$\%$ of those of the 7$\times$7 structures, and none of the unfeasible Fano resonances are enclosed by the convex-hull. To provide a visual perspective of the convex-hull, we repeat the algorithm in Fig.~\ref{fig:block diargram presented technique}(a) in two-dimensional (2D) and three-dimensional (3D) latent RS (dimensionality of the response RS being 2 and 3, respectively). We set 0.005 and 0.0035 as the autoencoder validation error (10$\%$ and 7$\%$ point to point error) for 2D and 3D spaces, respectively, while using 95$\%$ as the in-point percentage threshold for both spaces. Figures~\ref{fig:convexhull_results}(a) and \ref{fig:convexhull_results}(b) show the converged convex-hulls in 2D and 3D latent RSs; the calculated errors in testing the resulting convex-hulls are shown in Table~\ref{tab:inpoint_percentage}. It is clear from Table~\ref{tab:inpoint_percentage} that both 2D and 3D algorithms are capable of identifying the feasible responses with better than 99$\%$ accuracy, but their ability in identifying the unfeasible responses are reduced (from 0$\%$ to 10$\%$ and 35$\%$, respectively). In other words, by reducing the dimensions, it seems that the convex-hull covers a larger percentage of the overall area of the latent RS resulting in a larger error in identifying the unfeasible responses. 

\begin{table}
    \centering
        \caption{In-points percentage of each class of test patterns (14$\times$14 and 7$\times$7 responses as well as Fano line-shape resonances) lies in the 2-D, 3-D, and 6-D convex-hull as well as one-class SVM highest confidence region.}
    \label{tab:inpoint_percentage}
    \begin{tabular}{|c|c|c|c|}
    \hline
	 Algorithm  Class & Binary 14$\times$14 & Binary 7$\times$7 & Fano lineshapes\\
	\hline
	Convex 2-D & 99.2 \%  & 100 \% & 35 \%\\
	\hline
	Convex 3-D & 98.6 \%  & 99.8 \% & 10 \%\\
	\hline
		Convex 6-D & 91.8 \%  & 96\% & 0 \%\\
	\hline
    One-class SVM 2-D & 90.2 \%  & 90.6 \% & 0 \%\\
	\hline
    One-class SVM 3-D & 91.4 \%  & 89.4 \% & 0 \%\\
	\hline
	 One-class SVM 6-D & 88.2 \%  & 84.8 \% & 0 \%\\
	\hline
	
\end{tabular}

\end{table}

It is important to note that despite training with a non-aggressive success rate of 95$\%$, the convex-hull algorithm is capable of identifying all unfeasible responses as well as a large portion of the feasible responses. Nevertheless, the convex-hulls in Fig.~\ref{fig:convexhull_results} do not provide the level of feasibility or unfeasibility of a response. For example, it is not trivial to compare the robustness of the resulting designs for achieving two responses as there is not a simple one-to-one relation between the Euclidean distance to the convex-hull boundary and the feasibility of a response. To add this feature, we use the same training/validation data to train a one-class SVM to find the non-convex geometry of the feasible responses for the structure in Fig.~\ref{fig:nanostructures}(a) using 6D, 3D, and 2D latent RSs. While one-class SVM provides valuable information about the relative feasibility of each desired response, finding the optimum hyper-parameters (i.e., $\nu$ and $\gamma$) for one-class SVM is challenging. Here we use 500 validation patterns to cross validate the hyper-parameters and find  $\nu=0.4$ and $\gamma=4$ as the optimum parameters. Table~\ref{tab:inpoint_percentage} shows the results of testing the 6D, 3D, and 2D one-class SVM algorithms with the same data used for testing the convex-hull algorithm. Smaller success rates in identifying the feasible responses while perfect performance in identifying unfeasible responses are attributed to the tighter (non-convex) geometry of the one-class SVM. This is also seen from the graphical representation of the one-class SVM in the 2D latent RS in Fig.~\ref{fig:convexhull_results}(c). Note also that the absolute values of the success rates in Table~\ref{tab:inpoint_percentage} for one-class SVM depend on the definition of the highest confidence region. Reducing the level of confidence results in extension of its corresponding geometry and thus, a smaller error. In addition to the innermost geometry (also known as the highest confidence geometry) shown by the red curve in Fig.~\ref{fig:convexhull_results}(c), several boundaries are identified with different colors. Each added region corresponds to a different level of unfeasibility of a response that lies outside the highest confidence region. A quantitative measure for the level of feasibility of a response in this one-class SVM is the minimum Euclidean distance of that response form the boundaries of the highest confidence region. The calculated distance in the 6D one-class SVM for a series of responses of the structure in Fig.~\ref{fig:nanostructures}(a) are shown in Table~\ref{tab:distance_6-D}. The average distance for each class of responses in Table~\ref{tab:distance_6-D} is calculated over the entire set of those responses in the test dataset. In addition, for each class, a representative sample response and its actual distance from the geometry is shown.  A negative (positive) distance shows that the point lies outside (inside) the highest confidence region; the absolute value of the distance shows the relative unfeasibility (feasibility) of a response. Table~\ref{tab:distance_6-D} clearly shows that a smoother response (e.g., the first row of Table~\ref{tab:distance_6-D}) has a better feasibility than a sharper one (second row of Table~\ref{tab:distance_6-D}). It also confirms the unfeasibility of the ideal Fano and Lorentzian responses with Fano responses being farther from the feasibility region.

\section{Experimental Results}

\begin{figure}
\centering
\includegraphics[width=1\linewidth, trim={0cm 0cm 0cm 0cm},clip]{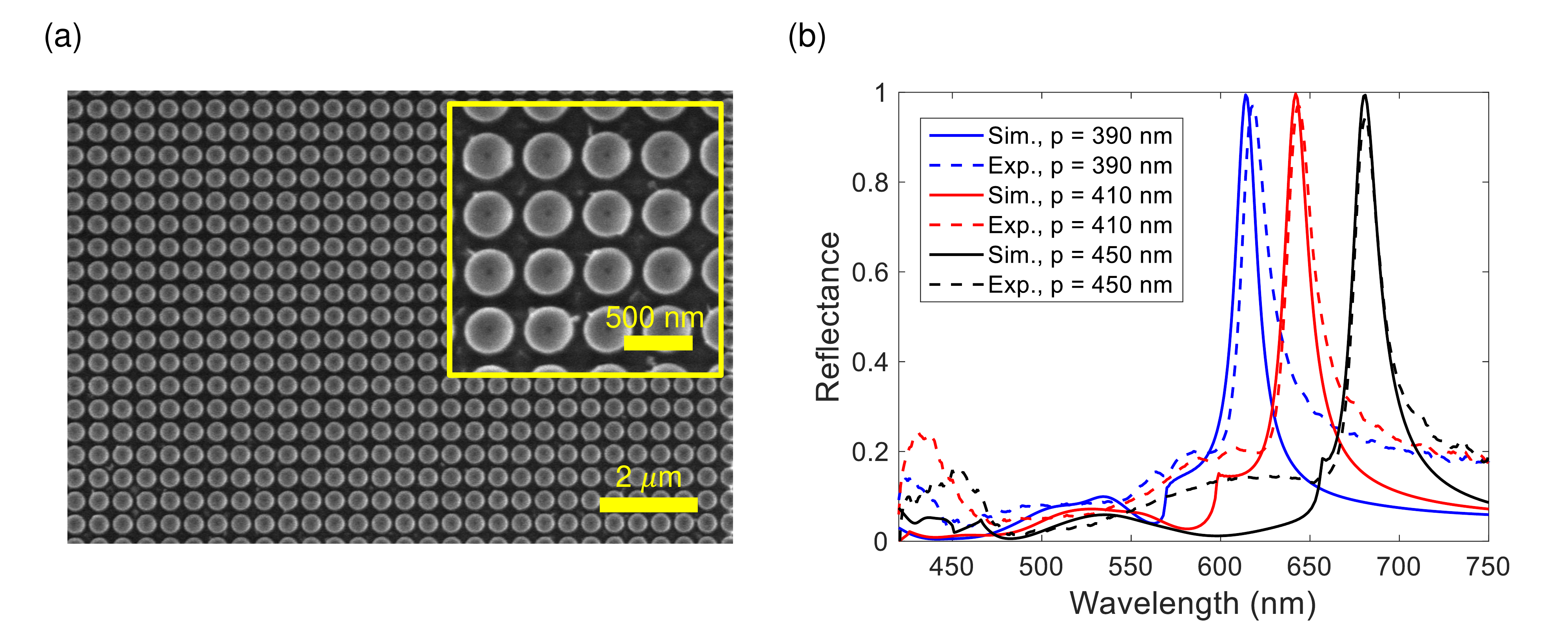}
\caption{(a) SEM images of a fabricated MS consisting of a rectangular lattice of cylindrical HfO$_2$ nanopillars with periodicity $p_x = p_y = p=450$ and radii $r_x = r_y = r=0.75p$ ($p_x, p_y,r_x,r_y$ are defined in \ref{fig:nanostructures}(c)). (b) The simulated and experimentally measured (Exp) reflectance spectra from MS (similar to the one in (a)) with periodicities $p$ = 390 nm, 410 nm and 450 nm and nanopillars with radii $r=0.75 \, p$.} 
\label{fig:HfO2_Sim_Exp}
\end{figure}

\begin{figure}
\centering
\subfigure[]{\includegraphics[width=0.49\linewidth, trim={0cm 0cm 0cm 0cm},clip]{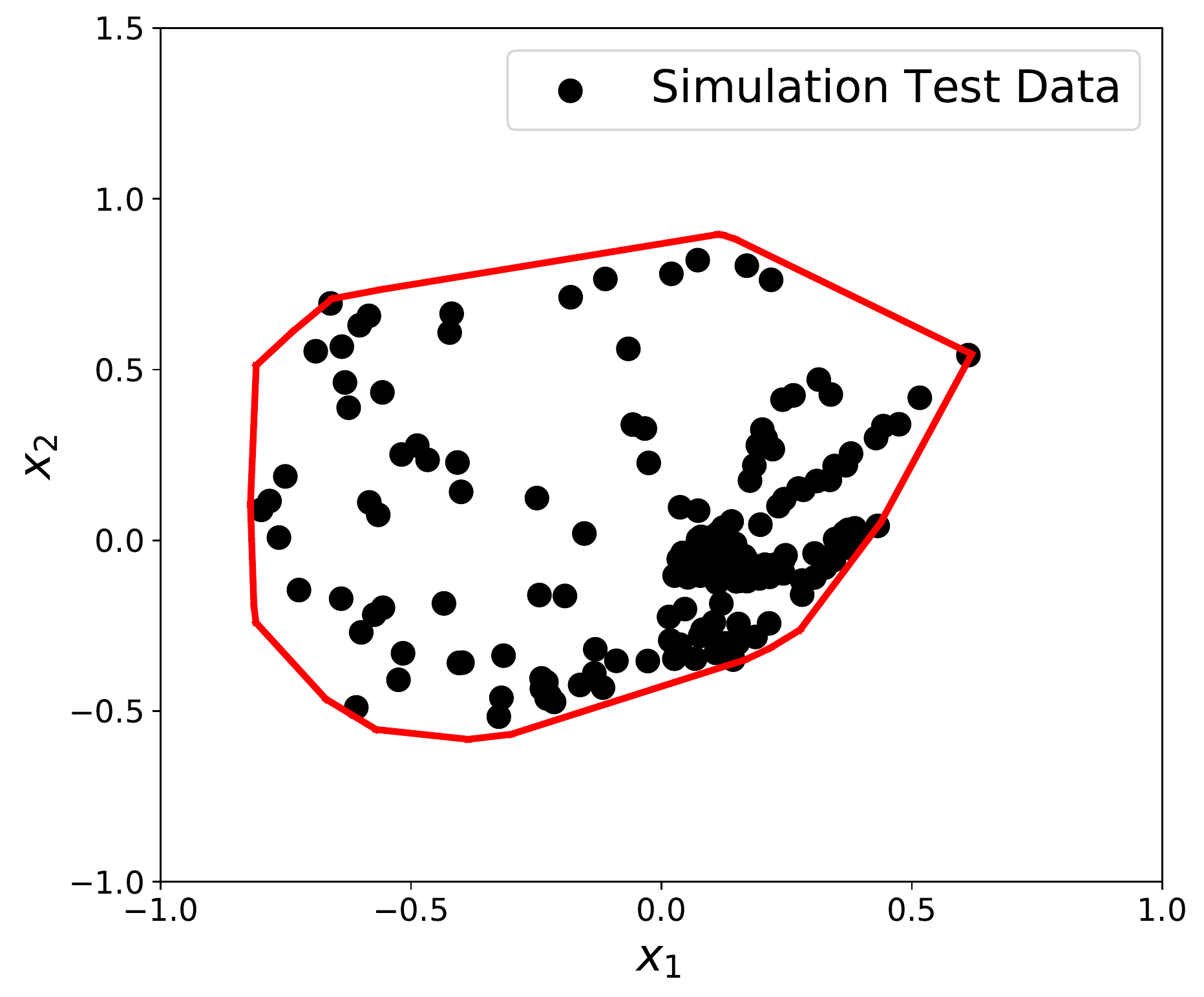}}
\subfigure[]{\includegraphics[width=0.49\linewidth, trim={0cm 0cm 0cm 0cm},clip]{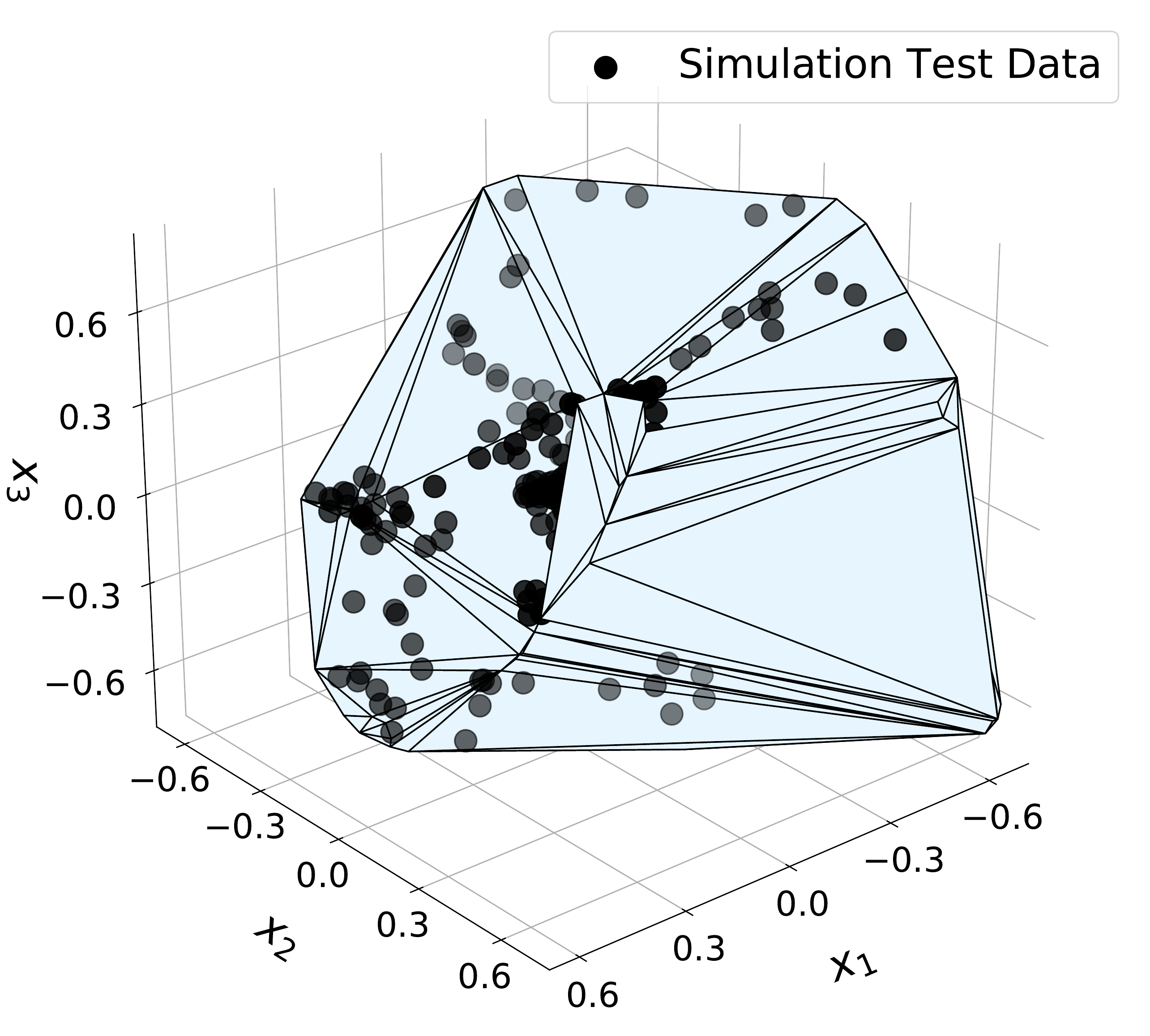}}
\subfigure[]{\includegraphics[width=0.49\linewidth, trim={0cm 0cm 0cm 0cm},clip]{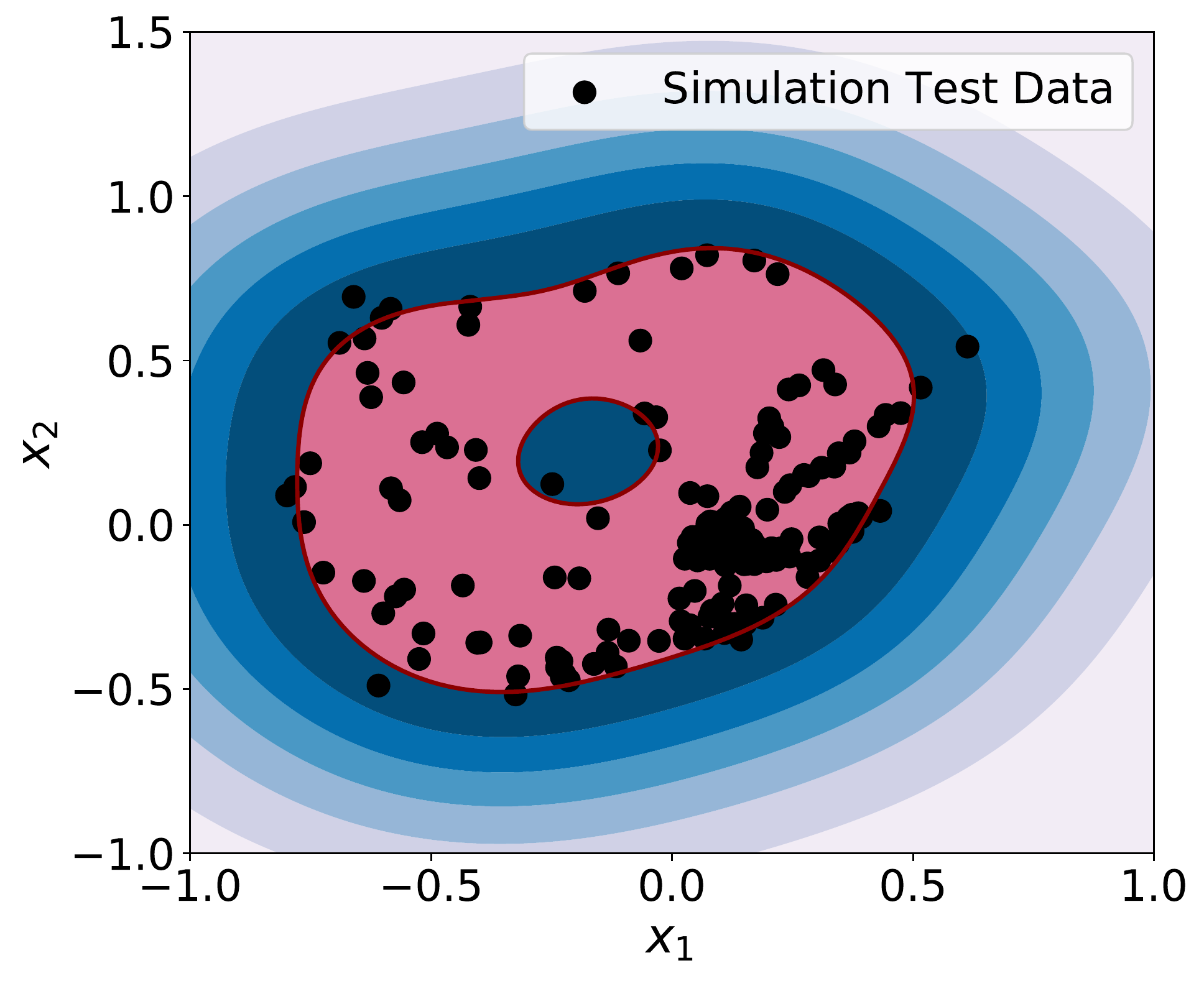}}

\caption{Representation of the (a) 2D convex-hull, (b) 3D convex-hull, and (c) one-class SVM for the HfO$_2$-based MS in Fig. \ref{fig:nanostructures}(c). The simulated structure consists of s square lattice of HfO$_2$ nanopillars with $p_x=p_y=p$ (ranging from 250 $nm$ to 450 $nm$) and radii $r_x=r_y=r$ ($0.6p<r<0.75p$). The feasible test patterns are also shown in each case demonstrating the capability of these algorithm in encompassing the feasible responses. }
\label{fig:simulation_nanopillar_conv_svm}
\end{figure}

To show the applicability of our technique in practical problems, without loss of generality, we choose a reflective structure formed by a low-loss dielectric (HfO$_2$) MS (\ref{fig:nanostructures}(c)), which can be experimentally fabricated and characterized. The training data for this structure is found by simulating the constituent unit cell with different geometrical parameters using FEM implemented in the COMSOL Multiphysics (see Methods section). The dimensions of the unit cell ($p_x$ and $p_y$ in Fig.~\ref{fig:nanostructures}(c)) can be changed between 250 nm to 450 nm while the radii of the nanopillars are proportionally modified ($r_{x}$ and $r_{y}$ in Fig.\ref{fig:nanostructures}(c)) from $r_{x,y} = 0.6 \,p_{x,y}$ to $r_{x,y} = 0.75 \,p_{x,y}$. The structure is illuminated by a TM-polarized plane wave of light at normal incidence, and the reflection coefficients at the far-field are calculated over the range of 400-800 nm wavelength range for 2400 patterns. The reflection spectra are uniformly sampled at 200 wavelengths to form a 200-dimensional RS. The resulting data is used to form the convex-hull and one-class SVM of the MS using the algorithms in Fig.~\ref{fig:block diargram presented technique}. The convex-hull-forming algorithm starts with an initial batch of data of 1000 patterns to train the autoencoder and form the 2D and 3D convex-hulls. In each iteration, we use 200 validation data for the autoencoder and 200 for the convex-hull. We select 5$\%$ and 95$\%$ as the validation thresholds for the autoencoder MSE and in-point percentages for the convex-hull, respectively. The algorithm converges after 5 (7) iterations for 2D (3D) RS space. After convergence, we test the algorithm using 200 ground-truth patterns whose results are shown in Figs.~\ref{fig:simulation_nanopillar_conv_svm}(a) and \ref{fig:simulation_nanopillar_conv_svm}(b). We also train a one-class SVM with $\nu=0.4$ and $\gamma=0.1$, and the results are depicted in Fig.~\ref{fig:simulation_nanopillar_conv_svm}(c). Our calculated in-points rate for the 2D (3D) convex-hull and the one-all SVM over the entire test data is 100$\%$ (98.5$\%$) and 93.5$\%$ (93$\%$), respectively. Table \ref{tab:nanopillars_average distance} compares the average distance of test response patterns from the boundary formed using one-class SVM for simulation and experiment. The results shows the algorithm is capable of providing a feasible geometry for experiment while it is trained on simulation data.

\begin{figure}
\centering
\subfigure[]{\includegraphics[width=0.49\linewidth, trim={0cm 0cm 0cm 0cm},clip]{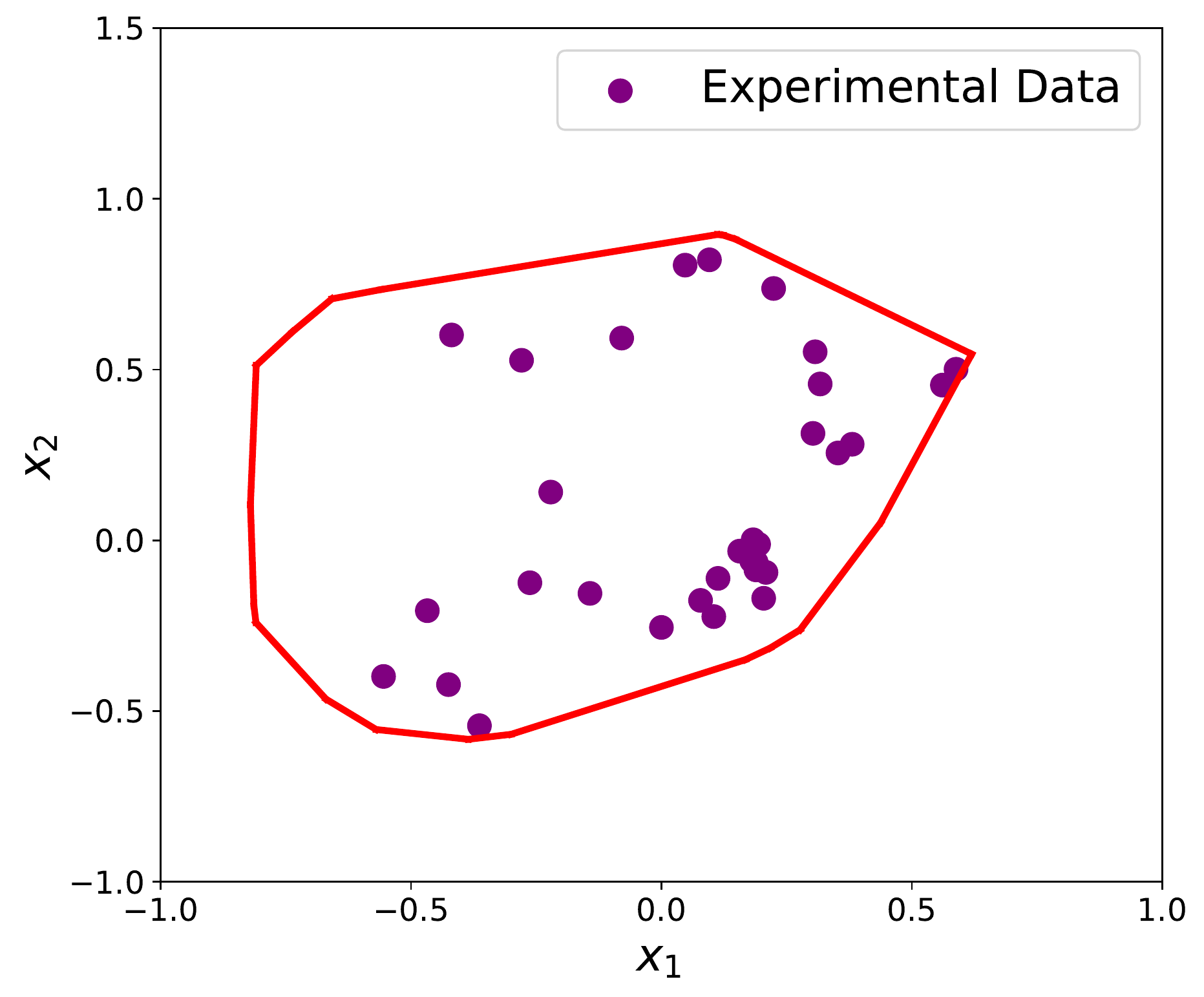}}
\subfigure[]{\includegraphics[width=0.49\linewidth, trim={0cm 0cm 0cm 0cm},clip]{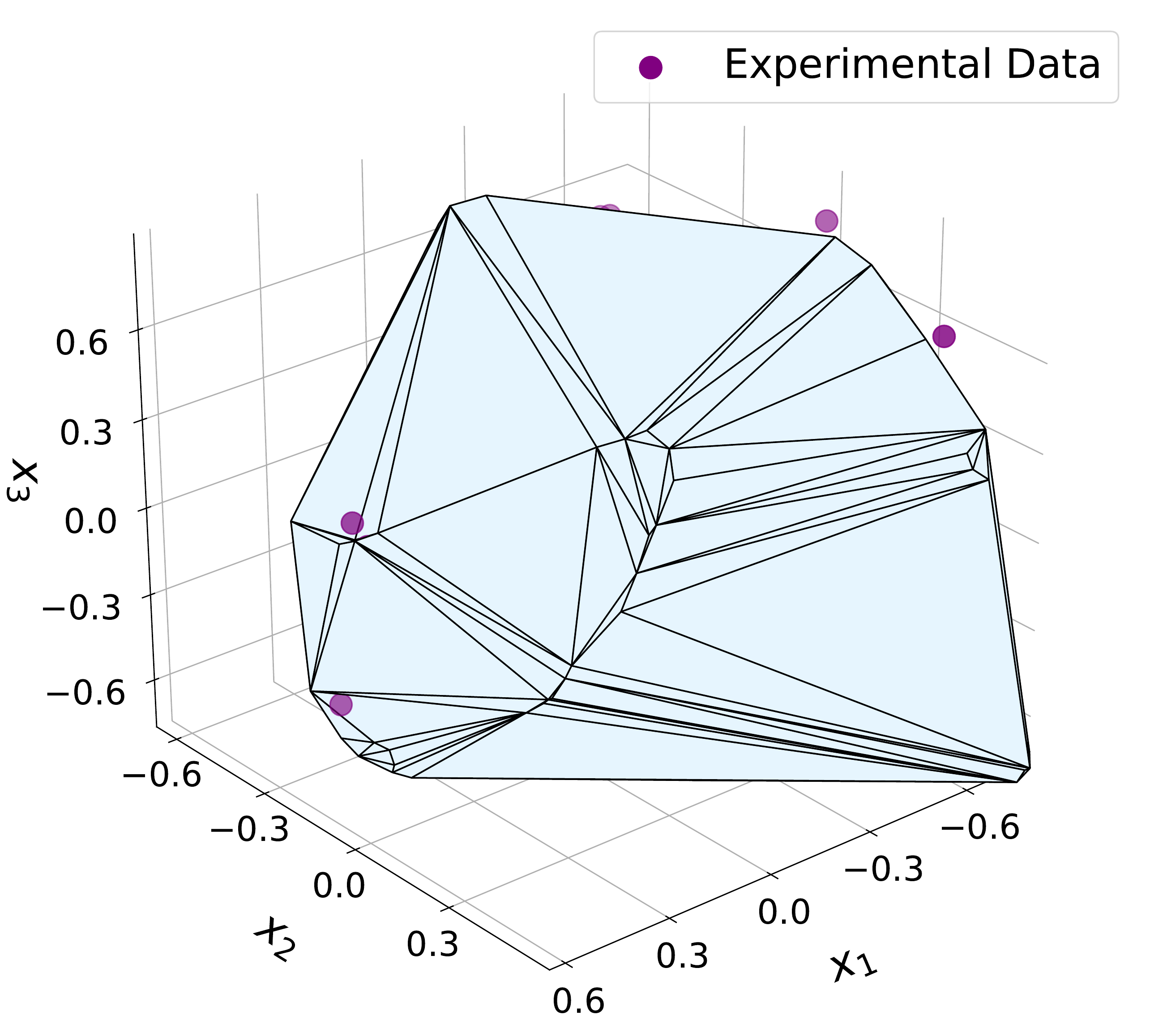}}
\subfigure[]{\includegraphics[width=0.49\linewidth, trim={0cm 0cm 0cm 0cm},clip]{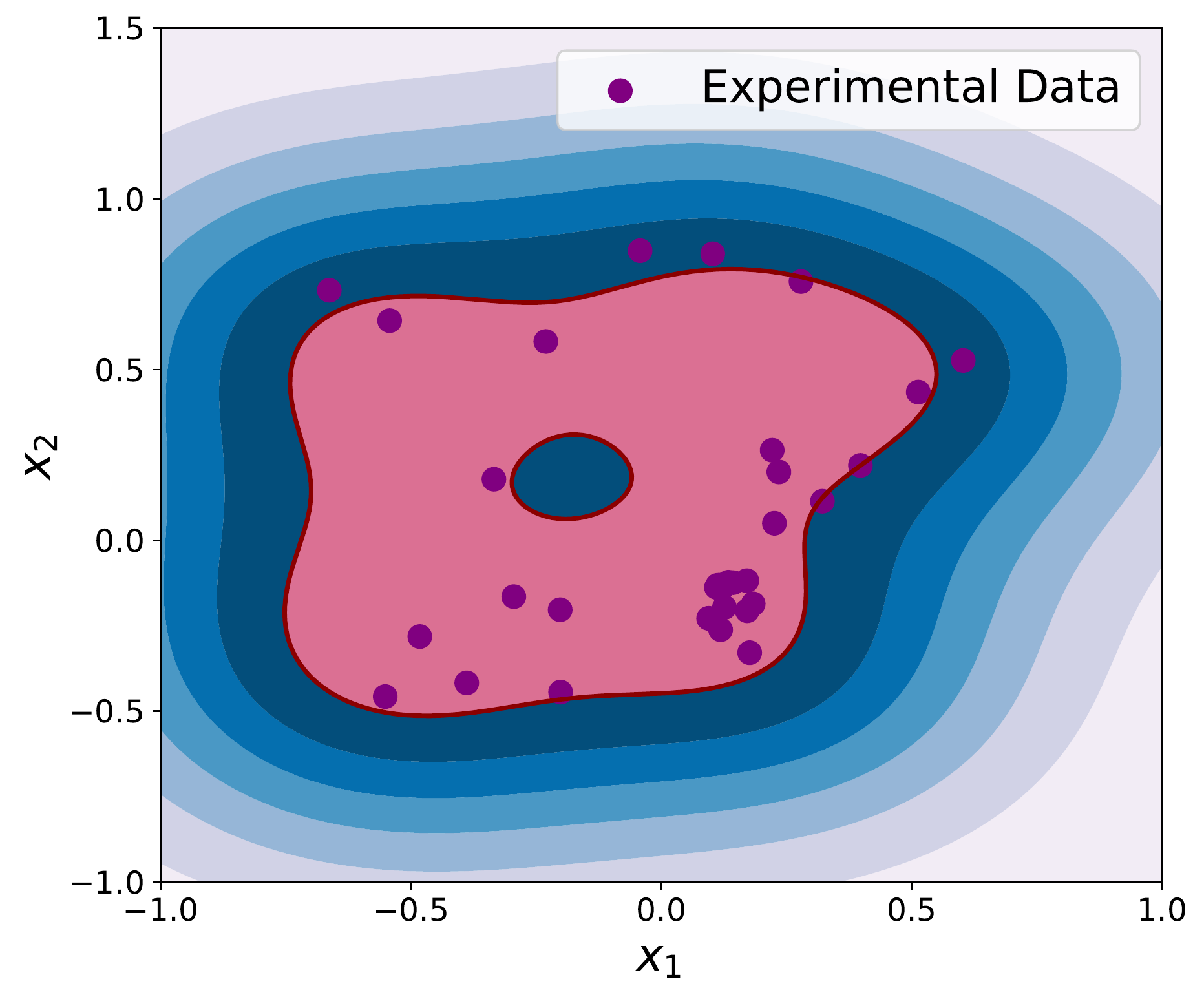}}

\caption{(a) 2D convex-hull, (b) 3D convex-hull, and (c) 2D One-class SVM for the dielectric MSs shown in Fig. \ref{fig:nanostructures} with properties described in the caption of Fig. \ref{fig:HfO2_Sim_Exp}. The experimentally measured data also shown. It is clear that almost all experimental results fall within the convex-hull or one-class SVM boundaries.  }
\label{fig:Fabrication_nanopillar_conv_svm}
\end{figure}

To evaluate the convex-hull experimentally, we fabricated several dielectric MSs with symmetric unit cells (i.e., $p_x = p_y = p$) with $250$ nm $<p<$ $450$ nm consisting of symmetric nanopillars (i.e., $r_x = r_y = r$) with $0.65 \, p <r< 0.75 \, p$ (see Methods for the fabrication details). The scanning electron microscopy (SEM) image for a fabricated MS with $p = 450$ nm and $r = 0.75p$ is shown in Fig.~\ref{fig:HfO2_Sim_Exp}(a),  Figure ~\ref{fig:HfO2_Sim_Exp}(b) shows a good agreement between the simulated and measured reflectance. Figure~\ref{fig:Fabrication_nanopillar_conv_svm} shows the placement of the experimentally measured responses in the RS space of the structure. It is clear that a large portion of the feasible responses fall within the convex-hull and the one-class SVM. In addition, the responses that fall outside the one-class SVM are close to the geometry of the highest confidence geometry with small distances. The calculated success rates of the 2D (3D) convex-hull and the one-class SVM in Fig.~\ref{fig:Fabrication_nanopillar_conv_svm} for the experimental results is 100$\%$ (87.87$\%$) and 87.87$\%$ (81.81$\%$), respectively, which is in good agreement with the theoretical results. Note that the despite using low dimensions for the latent RS, our techniques provide good success rates in identifying the feasible responses. 

\begin{table}
    \centering
        \caption{Average distance of test patterns for the responses achieved from nanopillars to the formed geometry of one-class SVM (using the training data achieved from the simulation patterns).}
    \label{tab:nanopillars_average distance}
    \begin{tabular}{|c|c|c|}
    \hline
	 Algorithm Type & Simulation & Experiment\\
	\hline

    One-class SVM 2-D & 3.655   & 3.657 \\
	\hline
    One-class SVM 3-D & 2.343   & 2.560 \\
	\hline

\end{tabular}

\end{table}

\section{Discussion}

The results in Figs. \ref{fig:convexhull_results}, \ref{fig:simulation_nanopillar_conv_svm}, \ref{fig:Fabrication_nanopillar_conv_svm}  and Tables \ref{tab:inpoint_percentage}, \ref{tab:nanopillars_average distance}, \ref{tab:distance_6-D} clearly show the power of GDL algorithms in assessing the feasibility of a desired response given a specific nanostructure design. They also show the advantage of one-class SVMs in providing a more quantitative measure for the level of feasibility of the desired response. This advantage comes from the fact that in one-class SVM, the geometric distance of a point in the latent RS from the boundaries of the one-class SVM is a good measure for the feasibility of the response while in convex-hulls, this relation does not hold necessarily. This advantage comes at the expense of more sophisticated training as the optimum hyper-parameters $\nu$ and $\gamma$ in SVM are not usually trivial to find. In practice, we first find the convex-hull of the feasible responses, and use it to find proper values of $\nu$ and $\gamma$ as explained in Supplementary Information section S2. Nevertheless, convex-hulls are helpful in providing the quick evaluation of the feasible response feasibility. The training process can also be simplified if more error is accepted. Note also that finding the exact geometry of the convex-hull and one-class SVM may not be important in design and optimization problems as the points on the boundaries correspond to less reliable responses that are prone to environmental changes or fabrication errors. We prefer the desired response to be in the middle of the one-class SVM. 

\begin{figure}
\centering
% \subfigure[]{\includegraphics[width=0.45\linewidth, trim={0cm 0cm 0cm 0cm},clip]{ConvHull_Train_2d.pdf}}
\subfigure[]{\includegraphics[width=0.49\linewidth, trim={0cm 0cm 0cm 0cm},clip]{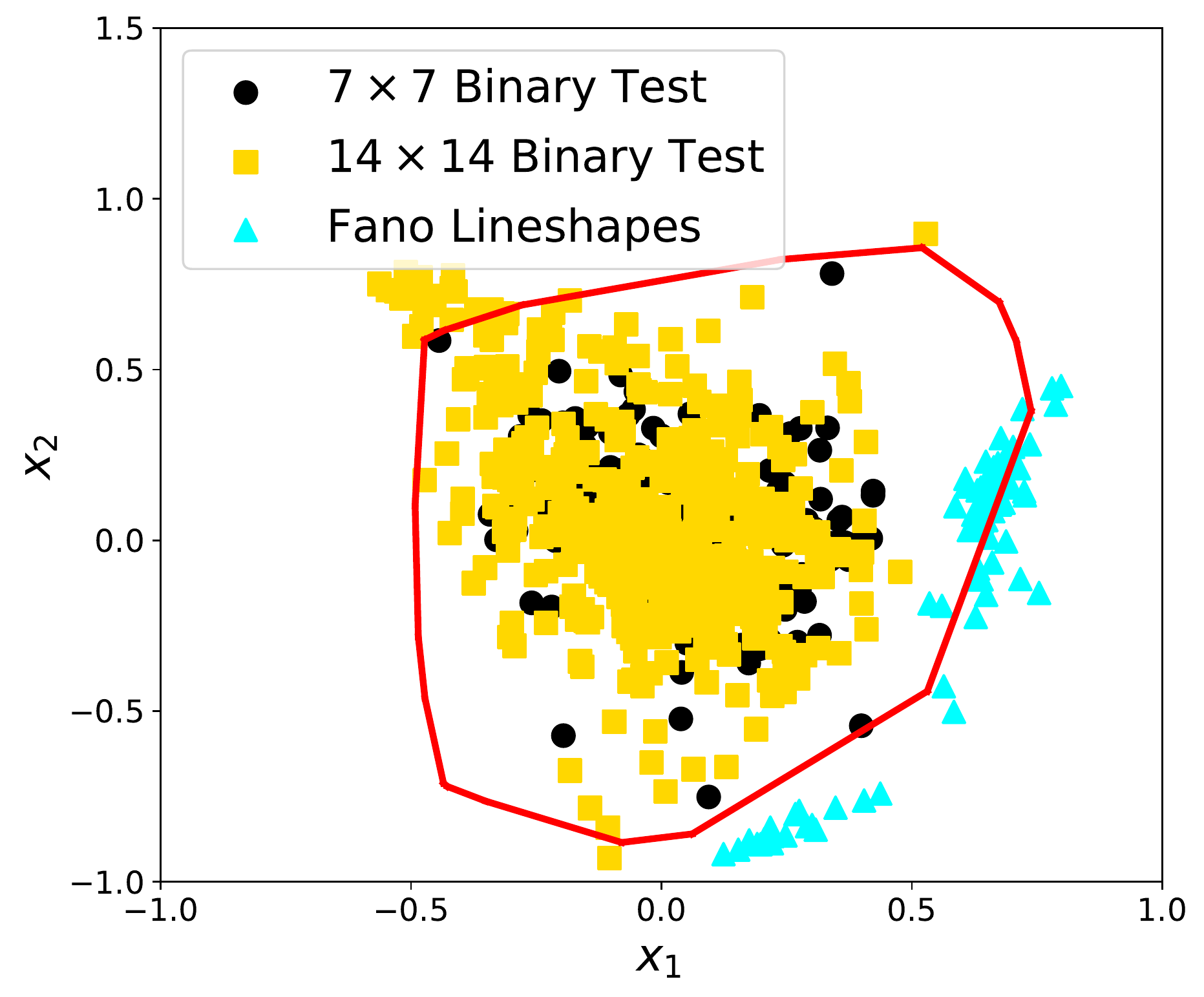}}
\subfigure[]{\includegraphics[width=0.49\linewidth, trim={0cm 0cm 0cm 0cm},clip]{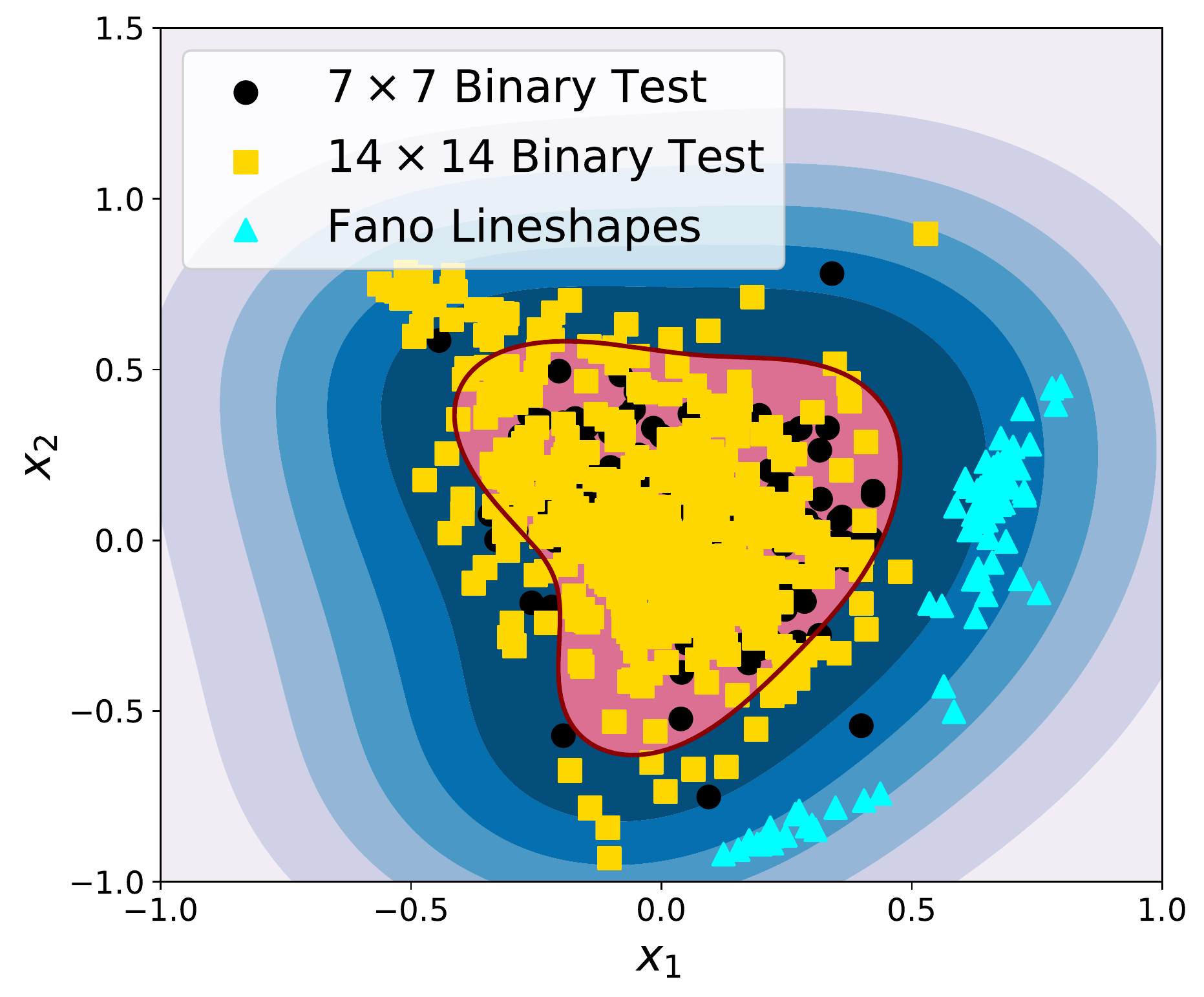}}
\caption{Representation of the convex-hulls in (a) 2D RS space of the 7$\times$7 binary structure in Fig.~\ref{fig:nanostructures}(b). The feasible responses for the 7$\times$7 and 14$\times$14 binary structures and the unfeasible idea Fano lineshapes are shown. (b) Non-convex geometry for the feasible responses found by one-class SVM algorithm along with feasible and unfeasible responses in the 2D latent RS for the 7$\times$7 binary structure in Fig.~\ref{fig:nanostructures}(b).}
\label{fig:conv7}
\end{figure}

In addition to the boundaries of convex-hull and one-class SVM in the latent RS, the area that is covered in that space by these shapes has important practical implications. The larger the area, the more capable the structure is in forming varieties of output responses. Figures~\ref{fig:conv7}(a) and \ref{fig:conv7}(b) show convex-hull and one-class SVM, respectively, in the 2D latent RS of the binary MS structure in Fig.~\ref{fig:nanostructures}(a) formed by 7$\times$7 array of nanostructures. For comparison, the responses used for the testing of the 14$\times$14 structure in Fig.~\ref{fig:nanostructures}(b) is also provided. For comparison Figs~\ref{fig:convexhull_results}(a) and (c) shows the 2D convex-hull and one-class SVM, respectively, for the 14$\times$14 structure while the testing data for the 7$\times$7 structure also presented. It is clear from Fig.~\ref{fig:convexhull_results} and Fig. \ref{fig:conv7} that the convex-hull and one-class SVM of the 7$\times$7 structure cover a smaller percentage of the 2D latent RS than those of the 14$\times$14 structure. This conclusion must be taken with the caveat that the latent RSs for the two structures are not necessarily the same.  Note that a wider range of responses may or may not be desirable for a design. For re configurable structures a wider response range is an advantage while for devices with a specific functionality wider response range usually is considered as unnecessary complexity of the selected structure. Figure~\ref{fig:convexhull_results} clearly shows that while technically none of the responses of the 7$\times$7 structure was used in training the convex-hull and one-class SVM of the 14$\times$14 structure, all these responses fall inside the convex-hull and one-class SVM as any 7$\times$7 structure can be formed using the 14$\times$14 structure. Figures~\ref{fig:conv7}(a) and \ref{fig:conv7}(b) also show that some of the responses achieved by the 14$\times$14 structure cannot be achieved using the 7$\times$7 structure while some of them can. This is an important observation as it confirms that using the 14$\times$14 structure for some responses might be unnecessary; the same response can be achieved by a much simpler structure (e.g.,7$\times$7 structure in this case) with less fabrication challenges and more robustness against fabrication imperfection. We believe this observation is an important potential application of convex-hull and one-class SVM in finding the most robust and least complex structures when starting from a non-optimal design. In addition, selecting a structure for which the desired response falls in the middle of the one-class SVM (i.e., has maximum distance from the boundaries) results in more tolerance against environmental changes and fabrication imperfections. 

The dimensionality reduction algorithm implemented by the autoencoder is an important step in reducing the required  computational resources for the convex-hull and one-class SVM. For any particular problem, the optimum dimension of the latent RS depends on the selection of the design and the redundancy of the response (i.e., the level of non-uniqueness). Thus, finding the optimum size of the latent RS is the initial step in implementing the algorithms of this paper. Once the size of the latent RS is selected, the required computation for the calculation of the convex-hull and one-class SVM are primarily for the training algorithm. In this paper, we mainly used the brute-force approach in starting with a training dataset and expanding it until the convex-hull (and subsequently the one-class SVM) pass the validation test. Further rigorous approaches muse be developed to minimize the computation required for training. One can also take advantage of the trade-off between the accuracy (or the error) and the computation requirement to optimize the training approach as explained above.   

Although the focus of this paper was the first demonstration of a GDL-based technique for studying the feasibility of a given response, this technique can be adopted for obtaining far more detailed information about the physics of nanostructures. As an example, Fig. \ref{fig:convexhull_results} clearly shows that the Fano-type resonances are clustered separately from the non-Fano-resonances. Further extension of this technique to separate more classes of responses (known as clustered homotops) is currently under investigation. 

\section{Conclusion}

In summary, we presented here a new approach to utilize AI for knowledge discovery in nanophotonics through training two well-known algorithms (convex-hull and one-class SVM). We showed that by combining the convex-hull (or one-class SVM) with DR by an autoencoder, we can find the range of feasible responses as well as the degree of feasibility of a desired response from any given class of EM nanostructure in its latent RS. By applying these techniques to a series of MSs, we showed the unique capabilities of one-class SVM and convex-hull in providing valuable insight about the capabilities of any EM nanostructure in providing different types of responses. While this is the first demonstration of an AI-based approach for such knowledge discovery, the presented techniques show great potentials in facilitating the understanding of the underlying physics of EM nanostructures as well as forming a more systematic approach in designing such nanostructures.

\section{Methods}

\subsection{Numerical Simulations}

All the simulations of the presented GDL method (including Quickhull, autoencoder, and one-class SVM) are implemented in Python using a simple personal computer with a 3.4 GHz core i7-6700 CPU and 16 GB of random access memory (RAM). The numerical simulations throughout the paper are carried out using the FEM implementation in COMSOL Multiphysics environment and interfaced to MATLAB to facilitate the process. For the design of single unit cells of any structure, periodic boundary conditions and perfectly match layers were considered in the lateral and vertical directions, respectively. A TM-polarized light in the range of 400-800 nm is launched into the simulation domain, and the co-polarized reflection coefficient is calculated at the location of the input port over the input bandwidth. The optical constants of Al, Al$_{2}$O$_{3}$ in Fig.~\ref{fig:nanostructures}(a) and (b) are obtained from Ref. ~{\cite{palik1998handbook}} using tabulated dielectric functions. The measured ellipsometry data for HfO$_{2}$ and quartz are used to simulate the structure in Fig.~\ref{fig:nanostructures}(c).

\subsection{Fabrication Process}

The dielectric MS shown in Fig.~\ref{fig:HfO2_Sim_Exp}(a) is fabricated on top of a quartz substrate. First, the substrate is cleaned and exposed to an oxygen plasma followed by spin-coating of a positive-tone electron-beam (e-beam) resist (ZEP-520A). The substrate was then soft-baked and coated with a conductive layer of Espacer to prevent charging effects during the e-beam writing process. Then, the sample is exposed to the e-beam (ELS-G100) to write the patterns followed by development in the diluted amyl acetate liquid. Atomic layer deposition of HfO$_{2}$ is performed using a standard two-pulse system of water and TEMAH at 90$^{\circ}$ under continuous flow of nitrogen carrier gas (Cambridge Nanotechnology). In the next step, the deposited top HfO$_{2}$ layer is etched using the inductively coupled plasma reactive ion etching process to reach the top surface of nanostructures. Finally, the sample is exposed to the ultraviolet light and oxygen plasma and soaked in the 1165 remover to remove the residue of the e-beam resist.

\section{Competing Interests}
The authors declare no competing interest.

 \begin{table}
        \centering
            \caption{Average distance of different classes of test data (14$\times$14 and 7$\times$7 responses as well as Fano and Lorenzian lineshape resonances) from the highest confident region border for one-class SVM. Distances for random samples represented in the most right column is also represented. The distances are calculated using Eq. S8.}
        \label{tab:distance_6-D}
        \begin{tabular}{|c|c|c|m{3.8cm}|}
        \hline
        & \mbox{\textbf{Average Distance}} &  \mbox{\textbf{Sample Distance}} &\quad \quad \mbox{\textbf{Sample Plot}} \\
    	\hline
    	\mbox{\textbf{Binary 14$\times$14}}  & \mbox{60.89} & \mbox{128.44} & \includegraphics[width=0.23\textwidth]{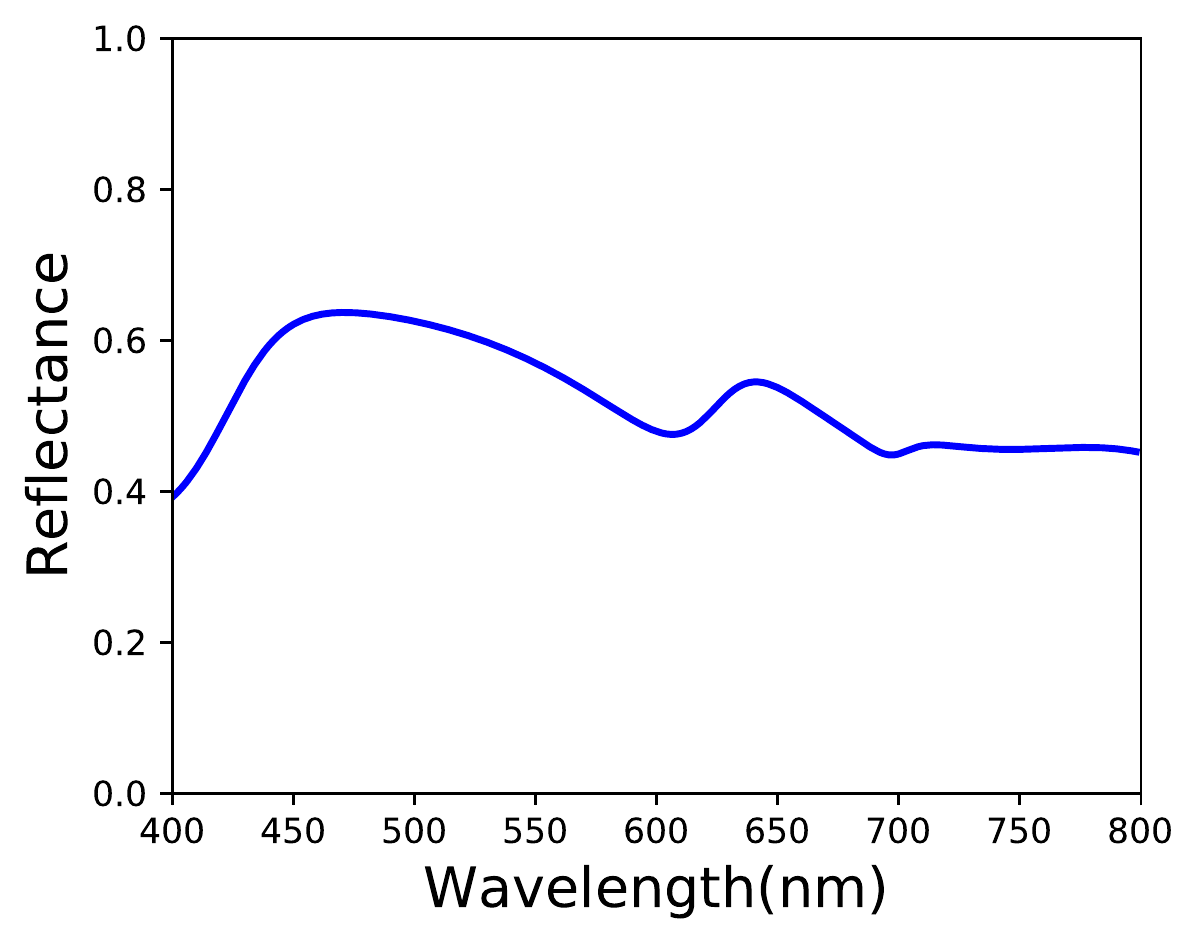}\\ 
    	\hline
    	\mbox{\textbf{Binary 7$\times$7}}  & \mbox{56.08} & \mbox{15.36}& \includegraphics[width=0.23\textwidth]{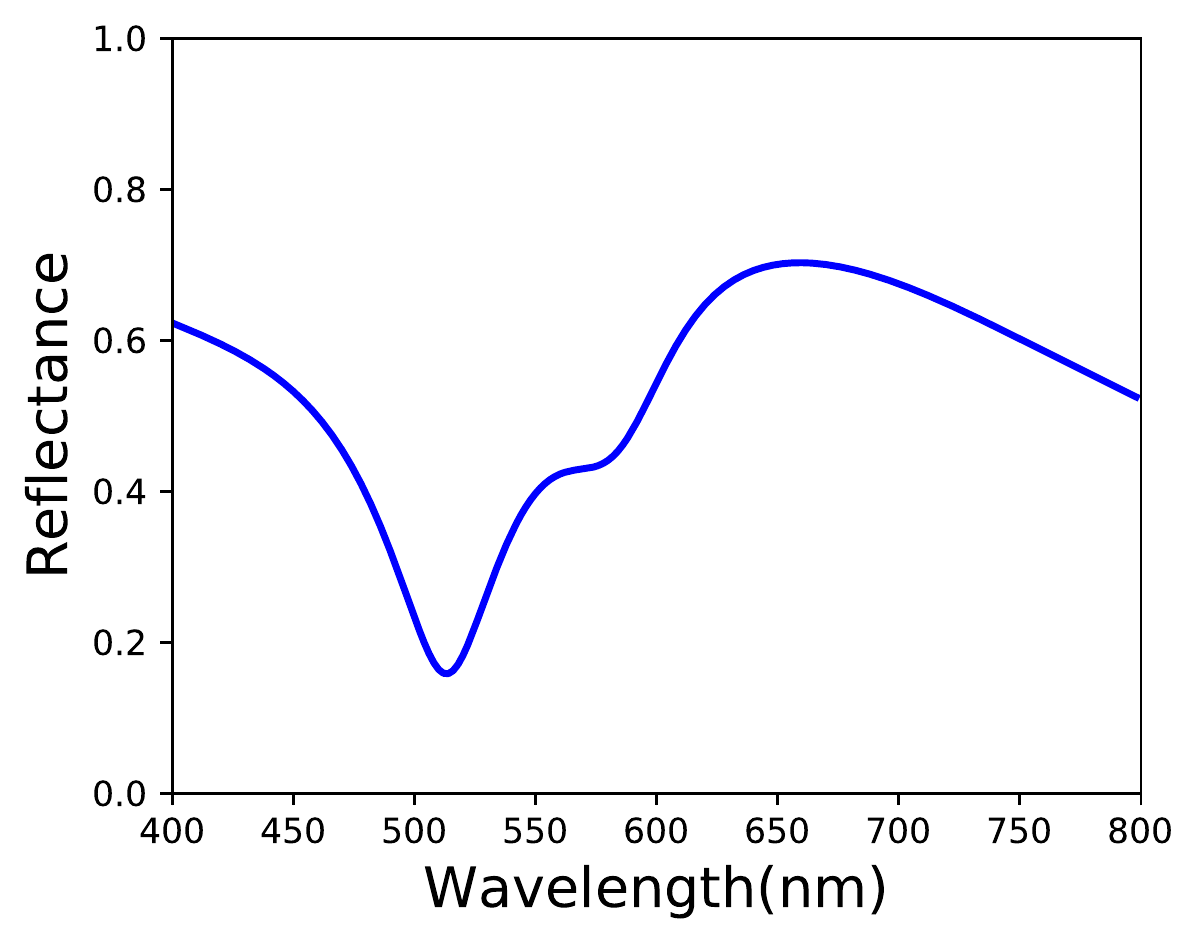}\\
    	\hline
        \mbox{\textbf{Lorentz Shape 1}}  & \mbox{-74.57} &\mbox{-62.53}& \includegraphics[width=0.23\textwidth]{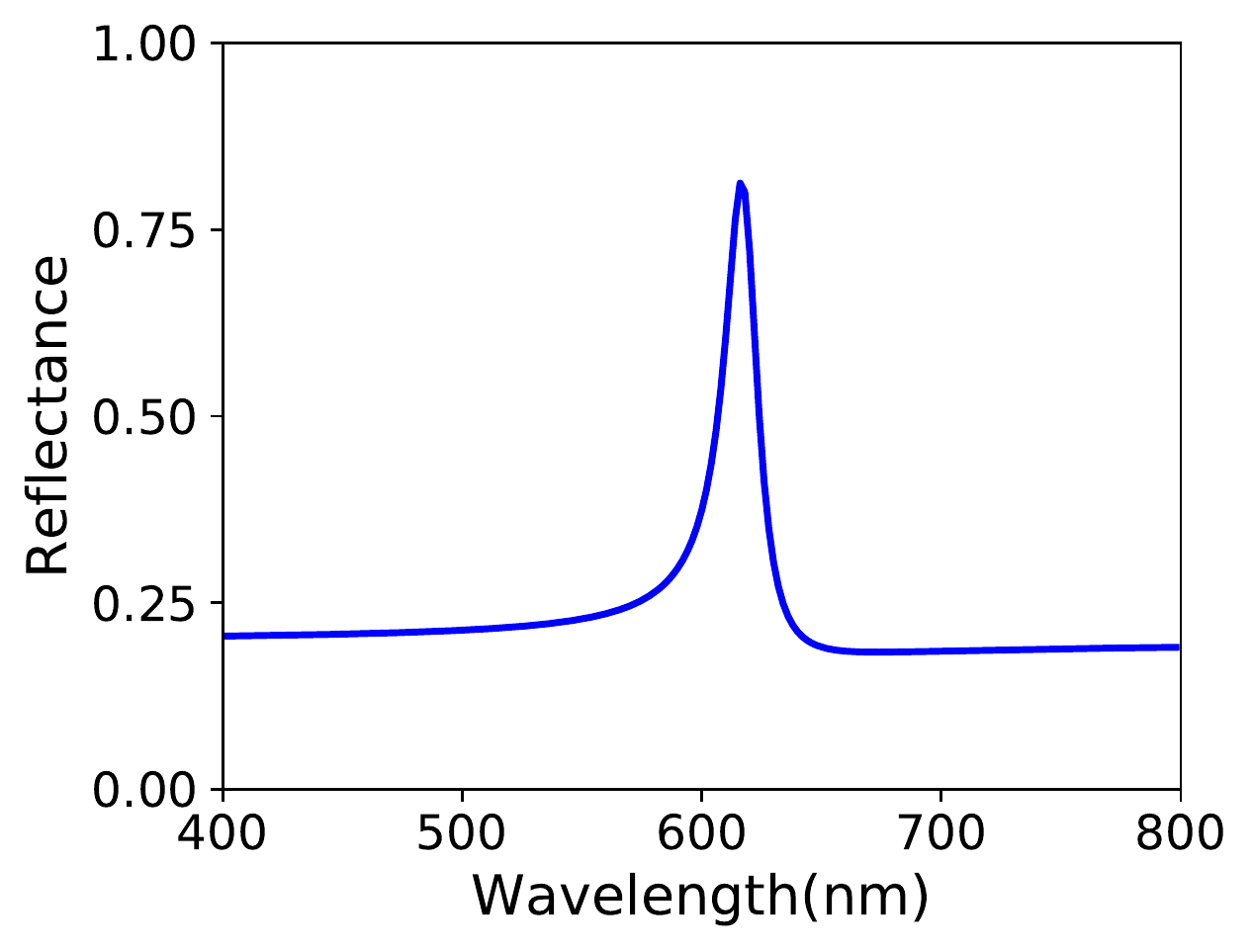}\\
    	\hline
        \mbox{\textbf{Lorentz Shape 2}}  & \mbox{-72.70} & \mbox{-46.75}& \includegraphics[width=0.23\textwidth]{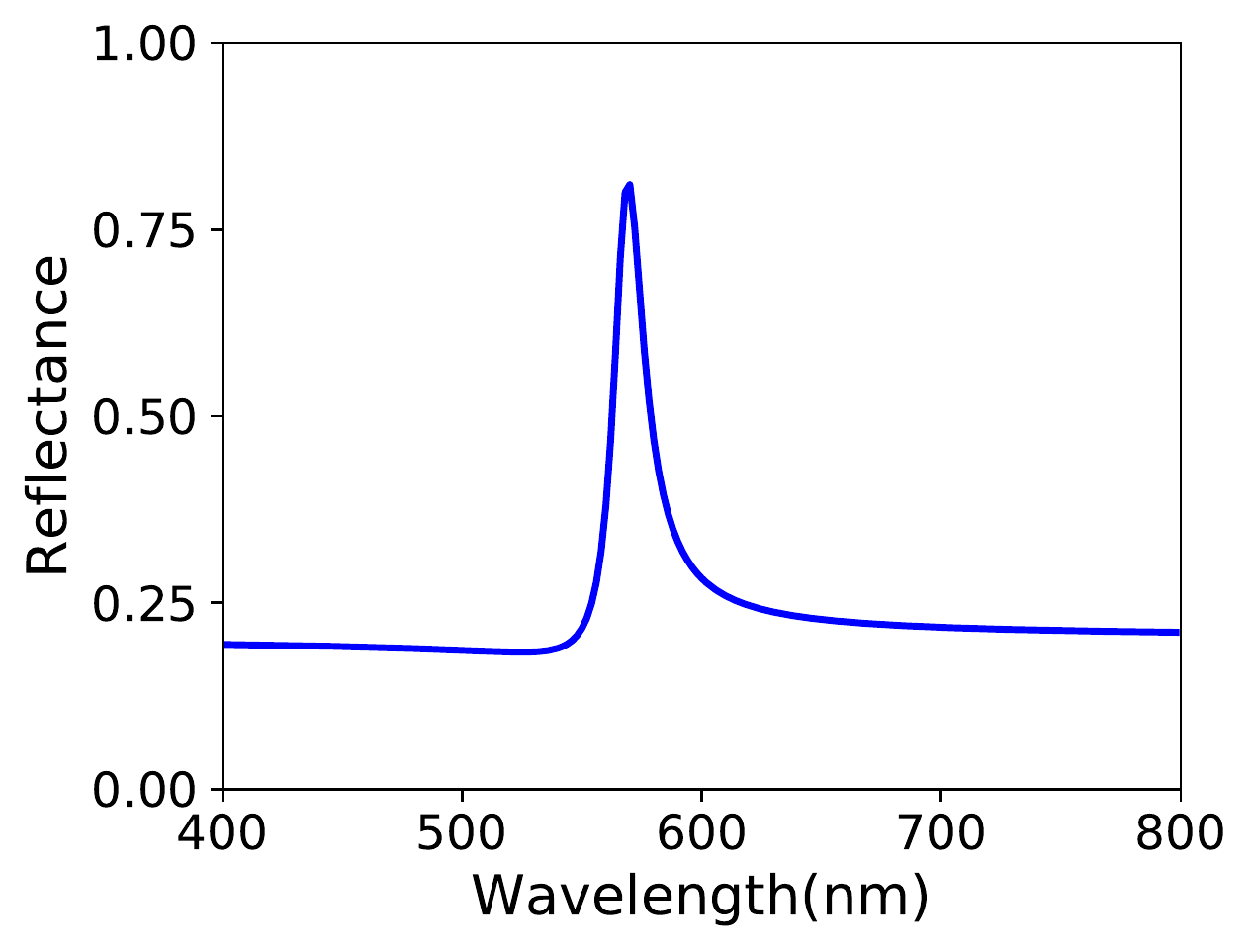}\\
    	\hline
        \mbox{\textbf{Fano Shape 1}}  & \mbox{-85.02} &\mbox{-60.71}& \includegraphics[width=0.23\textwidth]{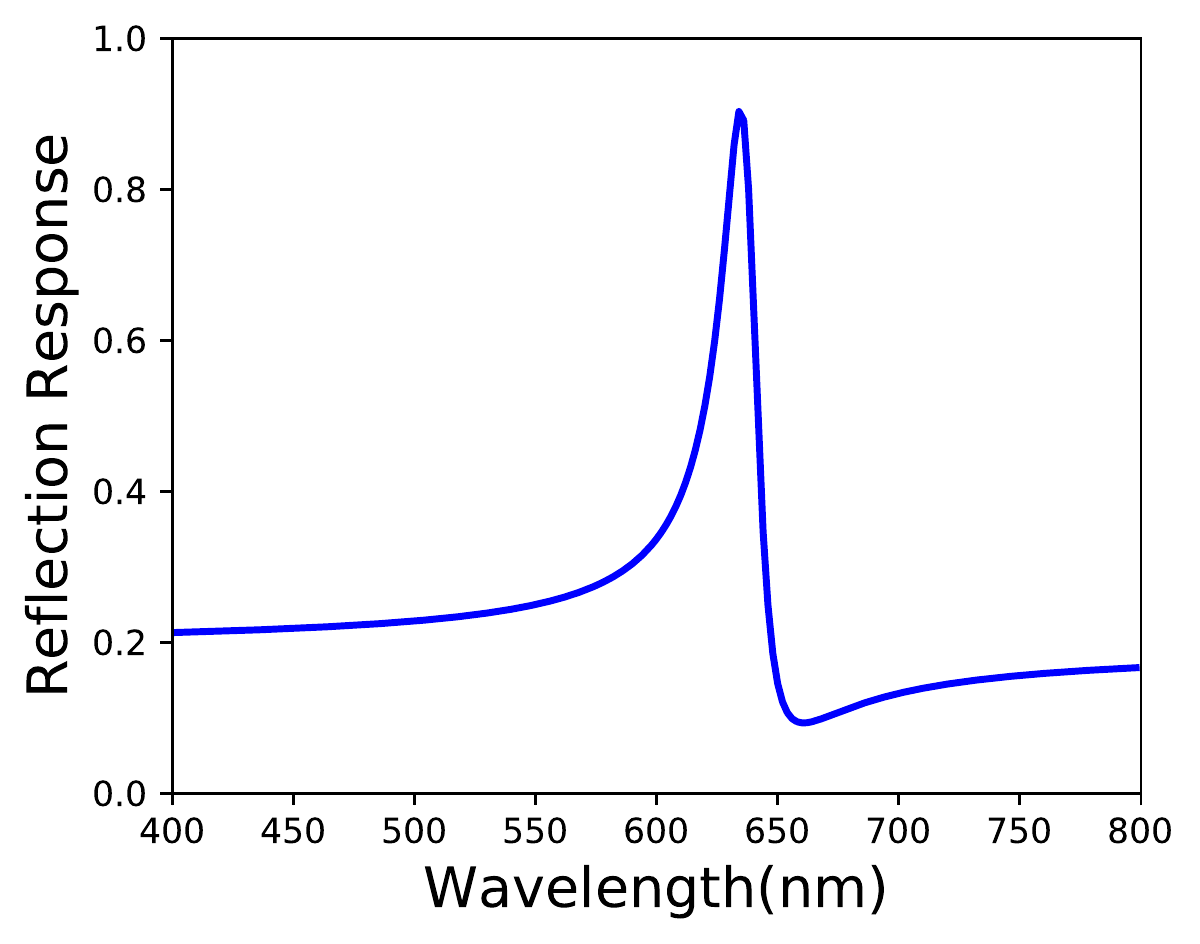}\\
    	\hline
    	\mbox{\textbf{Fano Shape 2}}  & \mbox{-80.39} & \mbox{-117.51}& \includegraphics[width=0.23\textwidth]{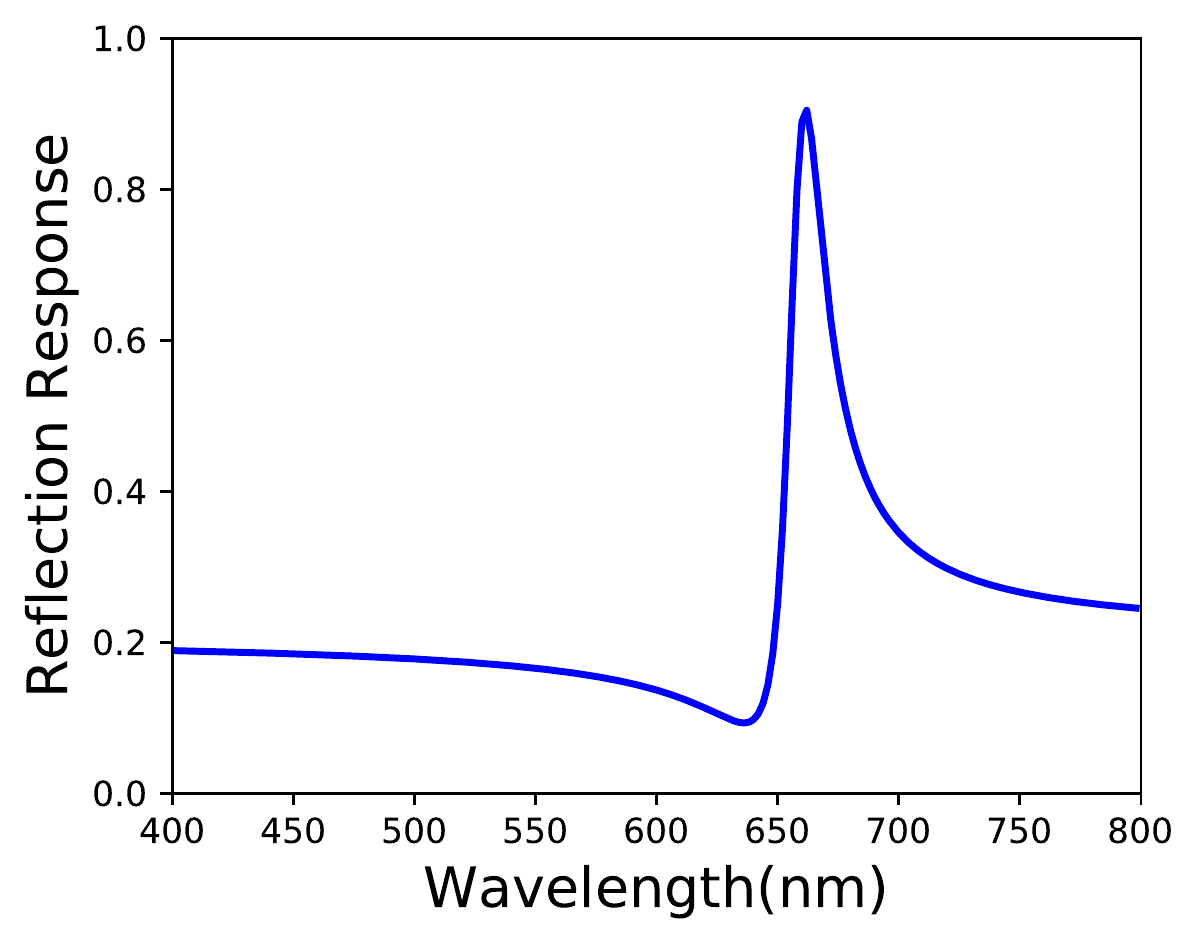} \\
    	\hline
    	
        \end{tabular}
    \end{table}

\clearpage
\newpage
\begin{centering}
\centering {\textbf{\Large{Supplementary Information}}}

\end{centering}
\section*{S1. Convexity and Convex-hull}

There are different ways to find a boundary that bounds a set of given points in space (e.g. Simplex, Voronoi Diagram, convex-hull, etc). The convex-hull of a set of points is the smallest convex set that contains all of them (see Fig. S1). Considering $x_1, x_2, \dots, x_k \in X$, the convex combinations of these points is defined as $\theta_1 x_1+\theta_2 x_2 + \dots + \theta_k x_k$ where $\theta_i \geq 0$ and $\theta_1+\theta_2+ \dots +\theta_k=1$. 
A set is convex (see Fig. S1) if and only if it contains all the convex combination of its points. The convex-hull of the set of points, $X$, is denoted as $conv \, X$ and is defined as

\begin{equation}\tag{S1}
    conv \, X=\{ \theta_1 x_1+\theta_2 x_2 +\dots+\theta_k x_k | x_i \in X ,\; \theta_i \geq 0,\;i=1,\, 2,\, \dots,\, k, \; \theta_1+\theta_2+\dots+\theta_k=1 \}
\end{equation}

The convex-hull operator on a set of points: is 1) extensive(i.e. the convex-hull of all sets in X is a superset of X), 2) non-decreasing (i.e. convex-hull of a subset of set X, is a subset of the convex-hull of X), and 3) idempotent (i.e. the convex-hull of the convex-hull of X is same as the convex-hull of X ). The convex-hull of any set of points is also a unique and closed set. 
\begin{figure}[b]
	\centering
	\includegraphics[trim=5cm 7cm 5cm 7cm,width=16cm,clip]{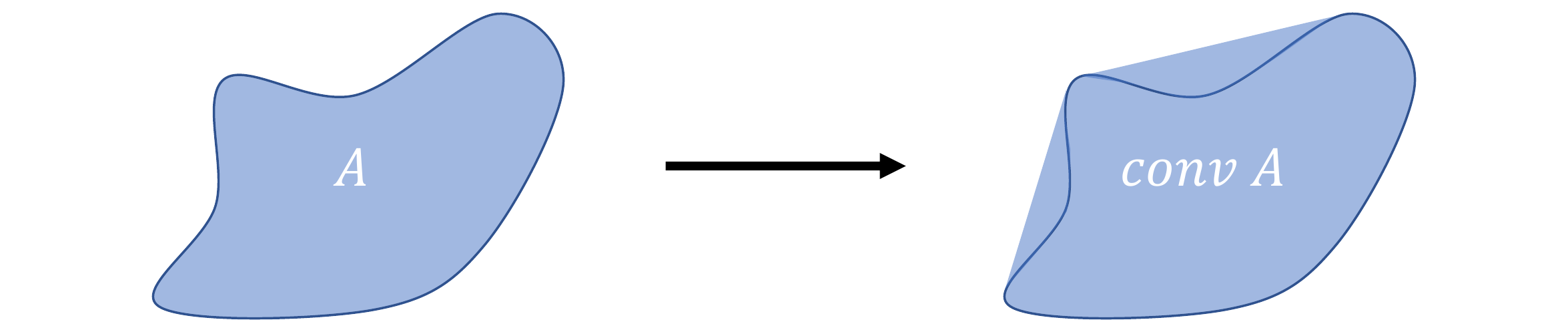}
	\caption*{Figure S1: Set $A$ shows a non-convex set of points. The convex-hull (i.e. $ conv \, A$) of this set is the smallest convex set that contains all the points in set $A$. }
	\label{fig:convexity}
\end{figure}

There are different algorithms presented in geometrical computation to form the convex-hull of a given set of points. One of the most effective and well-known algorithms is Quickhull. This algorithm finds the convex-hull of a set of points in $d$ dimensions using an effective method both in memory and computation. 
Given a set of $n$ data points with $r$ processed points, the algorithm is $O(n \log r)$ for $d\leq 3$ and is $O(nf_r/r )$ for $d>3$ ($f_r$ is the maximum number of facets for r vertices)\cite{barber1996quickhull}. The extreme points of a convex hull are referred as the vertices of the boundary of the convex hull. The running time of the algorithm depends on the number of facets and vertices of the convex-hull. Therefore, for sets with fewer extreme points it takes less time  for the algorithm to find the solution. 
A $d$-dimensional convex-hull can be shown using its vertices and $(d-1)$-dimensional facets. The ridges of the convex-hull are $(d-2)$-facets which are the intersection of the vertices in two neighboring facets. Quickhull forms the convex-hull using an incremental method based on Grunbaum's Beneath-Beyond theorem (see Fig. S2) as the following: 
\\
\textbf{Grunbaum's Beneath-Beyond Theorem:} Consider $H$ as the convex-hull of a set of points in $\textbf{R}^d$ and a point $p$ outside the convex-hull in $\textbf{R}^d-H$. $F$ is a facet of $conv(H \cup p)$ if and only if:\\
1) $F$ is a facet of $H$ and $p$ is below $F$, or\\
2) $F$ is not a facet of $H$, and its vertices are $p$ and the vertices of a ridge of $H$ that has one incident facet below $p$ and one above $p$.

\begin{figure}[t]
	\centering
	\includegraphics[trim=8cm 9cm 8cm 6cm,width=16cm,clip]{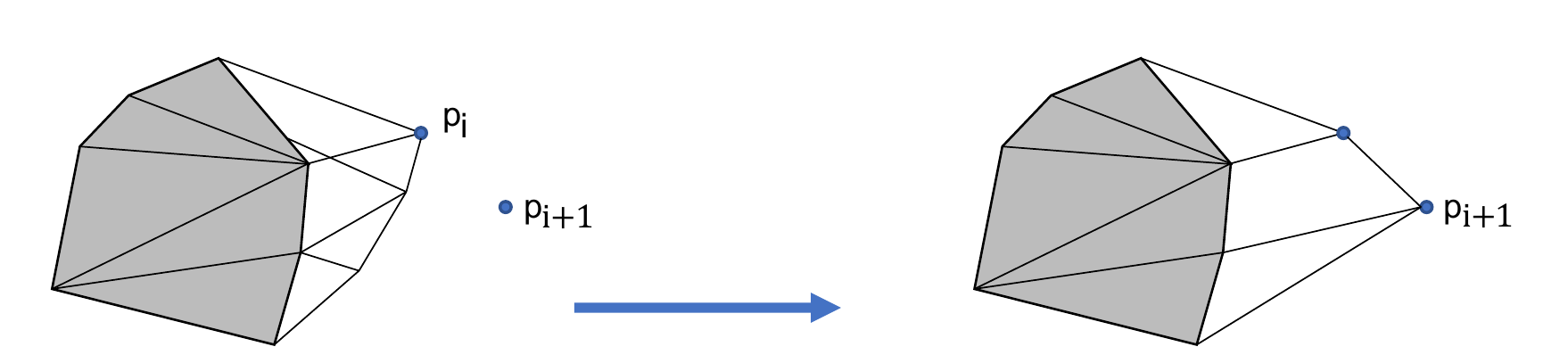}
	\caption*{Figure S2: Quickhull algorithm adds the farthest point in the outside set to the convex-hull at each iteration. The outside sets, the facets, ridges, and vertices will be updated in each step. This process continues until there is no outside point.}
	\label{fig:fig2_quickhull}
\end{figure}

The Quickhull algorithm starts with a set of points (i.e., a random subset of all training datapoints) and forms the initial convex-hull. All the points that lie outside of the initial convex-hull are considered as the outside set. The furthest point from the outside set is found at each iteration and based on Grunbaum's Beneath-Beyond Theorem, the facets, ridges, and vertices will be updated (see Fig. S2). This process will continue until convergence. The resulting convex-hull consists of all datapoints. 

After forming the convex-hull for a set $X$ in the latent space, we need to find out whether a given point $p$  lies inside the convex-hull or not. We first consider a random point $a$ outside of the convex-hull. We then connect $x$ and $a$ with a line segment $xa$ and find the number of its intersection with every vertex of the convex-hull. If the number of intersections is odd, the point lies inside the convex-hull. Otherwise, if the number of intersections is even or zero, this point is outside the convex-hull (see Fig. S3). 

\begin{figure}[t]
	\centering
	\includegraphics[trim=0cm 1cm 0cm 1cm,width=16cm,clip]{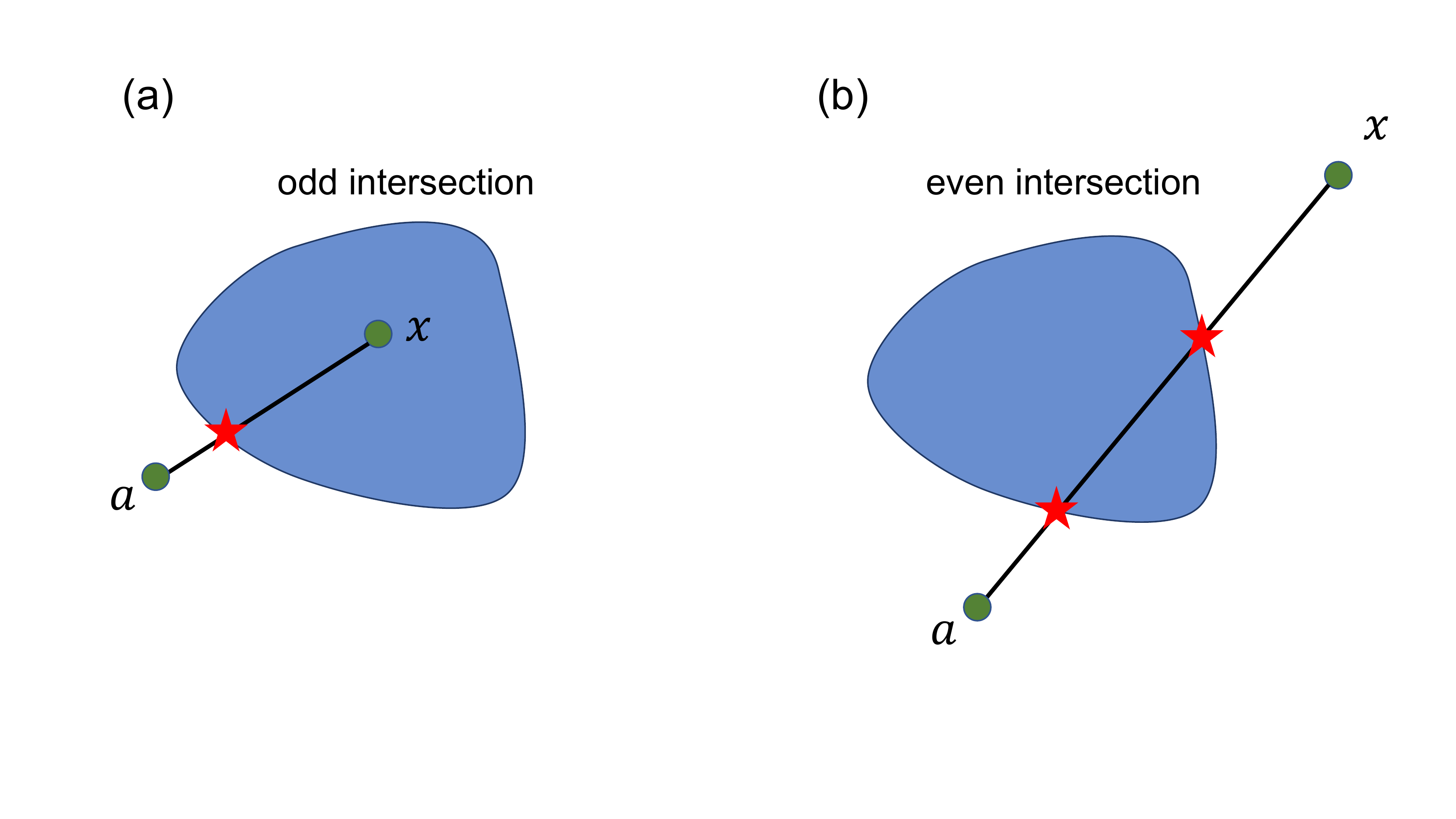}
	\caption*{Figure S3: The schematic of Inhull function for finding points inside and outside the convex-hull. To check if a sample point $x$ is inside or outside the convex-hull, the algorithm considers a random point $a$ outside of the convex-hull and finds the number of intersections of the line $xa$ with the convex-hull. If the number of intersections is odd, the point $x$ is inside (part (a)) and if it is even (part(b)), $x$ is outside the convex-hull.}
	\label{fig:fig2_quickhull}
\end{figure}

\section*{S2. One-class SVM}
As discussed before, the convex-hull just provides binary decisions about the feasibility or unfeasibilty of responses. To provide more information about the degree of feasibility one-class SVM (see Fig. S4) is used. 
Assume that the training data are $\mathbf{x}_1,\mathbf{x}_2, ..., \mathbf{x}_N \in \mathbf{X}$ where $N$ is the number of datapoints. Considering the mapping $\phi(\mathbf{x})$ from the feature space, $\mathbf{X}$, to a Dot Product space $\mathbf{F}$, the kernel function is defined as \cite{scholkopf1997kernel}:
\begin{equation}\tag{S2}
    k(\mathbf{x}_i,\mathbf{x}_j)=\langle \phi(\mathbf{x}_i),\phi(\mathbf{x}_j)\rangle
\end{equation}
There are different choices for the kernel function like Gaussian and polynomial kernel. 
In this research, we use the Gaussian kernel. 
\begin{equation}\tag{S3}
    k(\mathbf{x},\mathbf{y})=e^{-\frac{||\mathbf{x}-\mathbf{y}||^2_2}{\gamma}}
\end{equation}
One-class SVM can be formulated as an optimization problem which finds a hyperplane to separate datapoints in $\mathbf{X}$ from the origin in $\mathbf{F}$ and has the maximum distance from the origin \cite{scholkopf2000support}. This problem is formulated as a quadratic program:

\begin{equation}\tag{S4}
\begin{array}{rrclcl}
\displaystyle \min_{\mathbf{w}\in \mathbf{F},\xi \in \mathbf{R}^N, \rho \in \textbf{R}} & \multicolumn{3}{l}{\frac{1}{2}||\mathbf{w}||^2_2+\frac{1}{\nu N}\sum_{i=1}^{N}{\xi_i}-\rho}\\
\textrm{s.t.} & \langle \mathbf{w},\phi(\mathbf{x}_i)\rangle \geq \rho- \xi_i & \forall i \in \{1,...,N\}\\
&\xi_i\geq0    \\
\end{array}
\end{equation}
Here $\nu \in (0,1]$ is a free parameter of the algorithm. The slack variables $\xi_i$ let the algorithm to miss-classify some points to have a better generalization over unseen datapoints. Therefore, the free parameter $\nu$ penalizes the number of miss-classified points. For $\nu=0$, the penalty for the slack variables is infinite and the algorithm overfits to the training data while for larger $\nu$, more slack variables can have nonzero values and the algorithm under-fits. 
It is more practical to solve the dual problem for one-class SVM as \cite{scholkopf2000support}.

\begin{equation}\tag{S5}
\begin{centering}
  \begin{array}{rrclcl}
    \displaystyle \min_{\mathbf{\alpha}\in \mathbf{R}^N} & \multicolumn{3}{l}{\frac{1}{2}\sum_{ij} \alpha_i\alpha_j k(\textbf{x}_i,\textbf{x}_j)}\\
    \textrm{s.t.} & 0 \leq \alpha_i \leq \frac{1}{\nu N}& \forall i \in \{1,...,N\}\\
    &\sum_i \alpha_i=1  \\
\end{array} 
\end{centering}
\end{equation}
By solving this optimization problem, through a quadratic programming, the decision function becomes
\begin{equation}\tag{S6}
    f(\mathbf{x})=\sum_i\alpha_ik(\mathbf{x}_i, \mathbf{x})-\rho
\end{equation}
Here, $\rho$ can be recovered using the dual variables (i.e. $\alpha_i$). Those datapoints $\mathbf{x}_i$ with their corresponding optimized value $\alpha_i$ is nonzero are called support vectors. These datapoints are mainly close to the boundary and enforce the complexity of the boundary.  

\begin{figure}[t]
\centering
\includegraphics[width=1\linewidth, trim={7cm 5cm 7cm 5cm},clip]{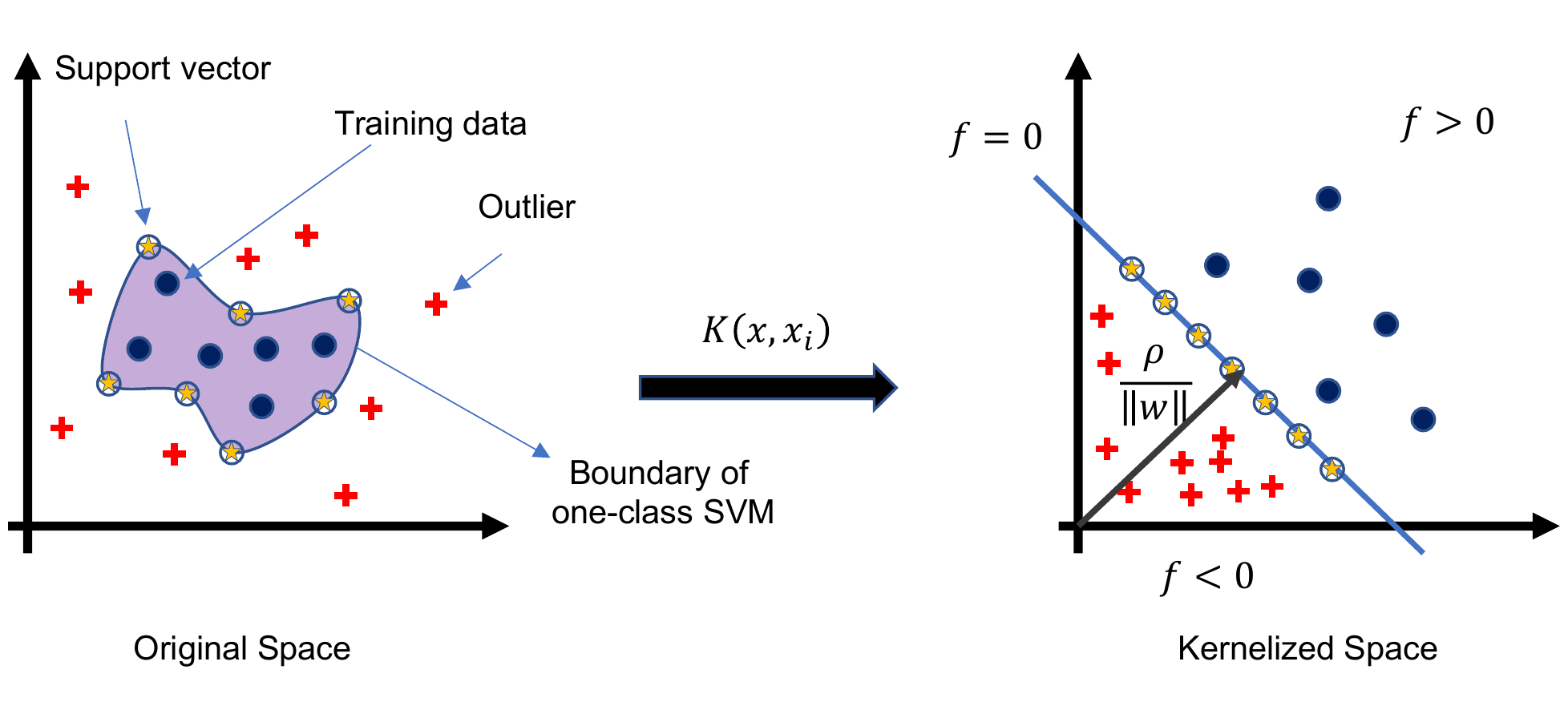}
\caption*{Figure S4: One-class SVM in the original space and kernelized space. All points are mapped to the kernelized space using a non-linear kernel functions, and the points are separated using a hyper-plane from the origin in the kernelized space, while in the kernelized space is linear. The decision boundary in the original space will be non-linear. }
\label{fig:MSE_Dimred}
\end{figure}

\section*{S3. Dimensionality Reduction}

The dimension of the original response space in the problem defined in this paper is 200. Defining convex-hull and one-class SVM in a high-dimensional space faces two major issues. 1) The distances and patterns in a high-dimensional space cannot be easily interpreted resulting in low performance, and 2) running time of the Quickhull algorithm increases as the dimensionality increase resulting in impractical computing for a high-dimensional space. To address these problems we used auto-encoder to reduce the dimensionality of the response space. The optimum dimensionality of the latent space (or the reduced RS) is find by minimizing the MSE in training the autoencoder. In this work, we use trial-and-error to find the optimum dimensionality of the response space. Figure S6 shows the MSE for DR of the structure in Fig.2 (b) as a function of the dimensionality of the reduced RS for an autoencoder with 7 layers (200,100,50,$d$,50, 100, and 200 nodes with $d$ being the dimensionality of the reduced RS). Based on Fig. S5 we choose the dimensionality of the reduced RS to be 6. In addition, we use 2D and 3D reduced RS for a better visualization. Further details for the implementation of the autoencoder is presented in Table S1.  

To have a better sense of the efficiency of the algorithm, we define point-to-point error ($Error_p$). Assume that we have $n$ response patterns and each response pattern achieved by discretizing the the response by measuring reflectance ($r$ and $\hat{r}$ represent ground truth and estimated reflectance respectively) in $m$ different wavelengths (i.e $\lambda$). The point-to-point error becomes:

\begin{equation}\tag{S7}
    Error_{p}=1/mn\sum_j^n\sum_i^m\frac{|\mathbf{r_i(\lambda_j)}-\mathbf{\hat{r_i}(\lambda_j)|}}{\mathbf{|r_i(\lambda_j)|}}
\end{equation}

\section*{S4.  Results for Plasmonic Oligomer}

Figure S7 (a) shows the results for one-class SVM in 2D of a 14$\times$14 array (see Fig. 2 (b)) of plasmonic structure. As it is shown, this structure has some sharp resonances, which are not likely for the $14 \times 14$ binary structure in fig. 2 (b). Considering the physical properties of the $14 \times 14$ binary structure, the responses with sharper resonances (e.g. responses 1, 2, 12, 10 in Fig. S7 (b)) have more distance to the feasible region and are less likely. On the other hand, the smoother responses (e.g. responses  7, 8, 9 in Fig. S7 (b)) that do not have sharp Fano-type resonances are more likely and have less distance to the feasible region.  

\begin{figure}[t]
\centering
\includegraphics[width=0.5\linewidth, trim={0cm 0cm 0cm 0cm},clip]{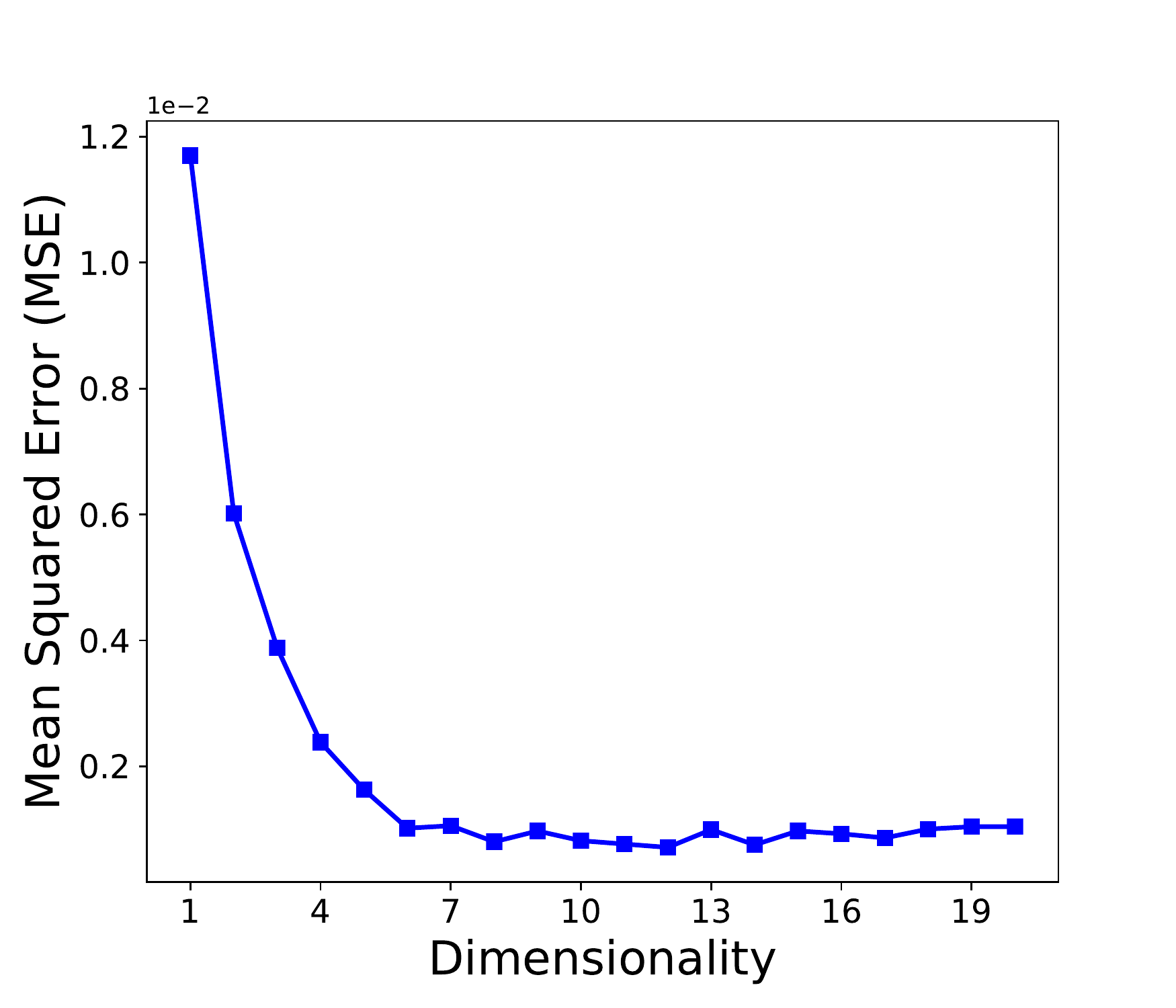}
\caption*{Figure S5: MSE for dimensionality reduction for the responses of the $14\times 14$ binary structure in Fig. 2(b) using an autoencoder.}
\label{fig:MSE_Dimred}
\end{figure}

\section*{S5. Fano-lineshapes}
To understand the capabilities and limitations of the binary structures, we tested the algorithm with Fano lineshapes. These type of resonances can be observed in the reflectance response of the all-dielectric MS consisting of HfO$_2$ nanopillars (NPs) shown in Fig. 2(c) in the main text, or in the scattering response from the plasmonics oligomers shown in Fig. S5(a-c). In the former case, the reason for the appearance of sharp Fano resonances are the strong coupling between the directly reflected light and the local magnetic dipole mode inside the NPs. For the latter case, the destructive interference between two sub-radiant and super-radiant modes supported by the nanoclusters result in a dip in the scattering spectrum at the Fano frequency \cite{king2015fano}. Here, to introduce these types of Fano resonances to our algorithm, we use the following standard formula for the reflectivity R of an arbitrary radian frequency $\omega$ :

\begin{equation}\tag{S8}
    R = a + (b+ic) \frac{\gamma}{i(\omega - \omega_0) + \gamma}
\end{equation}

% \cite{shen2015structural}
where $a$, $b$, and $c$ are the constant real numbers, $\omega_0$ is the central radiant resonant frequency, and $\gamma$ is the overall damping rate of the resonance. The quality factor ($Q$) of the fano resonances is calculated by $Q=\omega_0/\gamma$. Figure S6(d) represents different types of Fano lineshapes in the reflection spectrum from an all-dielectric MS consisting of HfO$_2$ NPs (Fig. 2(c)) and three plasmonic oligomers (Fig. S6(a-c)).

% \begin{figure}
% \centering
% \subfigure[]{\includegraphics[width=0.7\linewidth, trim={0cm 0cm 0cm 0cm},clip]{Experiment_SVM_nums.pdf}}
% \subfigure[]{\includegraphics[width=0.5\linewidth, trim={0cm 0cm 0cm 0cm},clip]{Experimental_plots.pdf}}

% \caption{(a) The trained one-class SVM using nanopillar structures responses (simulation data) and representation of the response instances achieved from the experiment. (b) Representation of the 20 random responses achieved from oligomer nanostructures with 20 set of random design parameters. Corresponding Oligomer reflectance to each point represented in part (a) by the purple dots.}
% \label{fig:zoom in nanopillars_nanopillar_conv_svm}
% \end{figure}

\begin{figure}[t]
	\centering
    \includegraphics[trim=0cm 0cm 0cm 0cm,width=1\textwidth,clip]{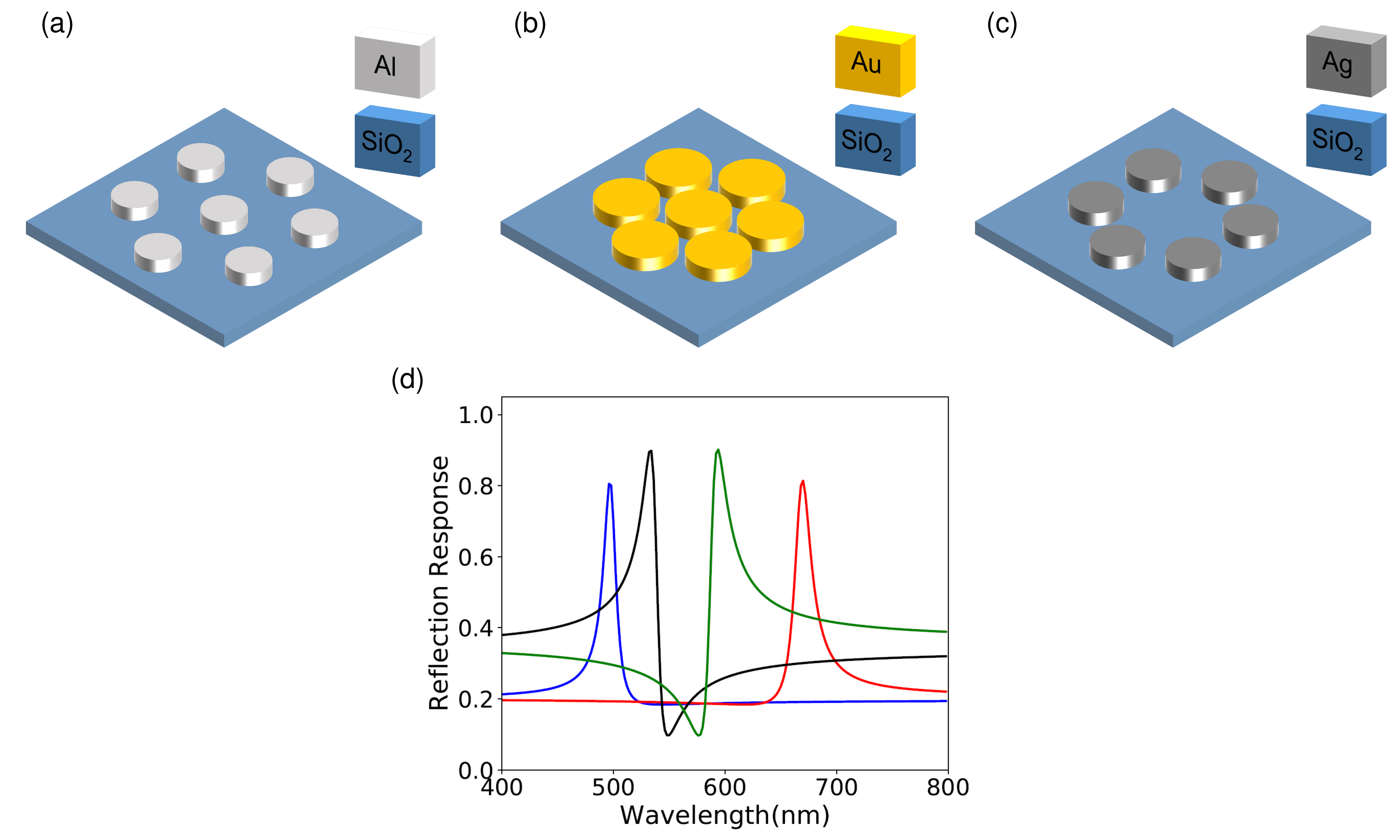}
	\caption*{Figure S6: Schematics of plasmonic oligomer made of (a) aluminum (Al), (b) gold (Au), and (c) silver(Ag) nanoclusters. (d) Ideal Fano lineshapes used for testing the capabilities of binary structure in Fig. 2(b) in achieving sharp responses.}
	\label{fig:Oligomer_str}
\end{figure}

\begin{figure}
\centering
\subfigure[]{\includegraphics[width=0.7\linewidth, trim={0cm 0cm 0cm 0cm},clip]{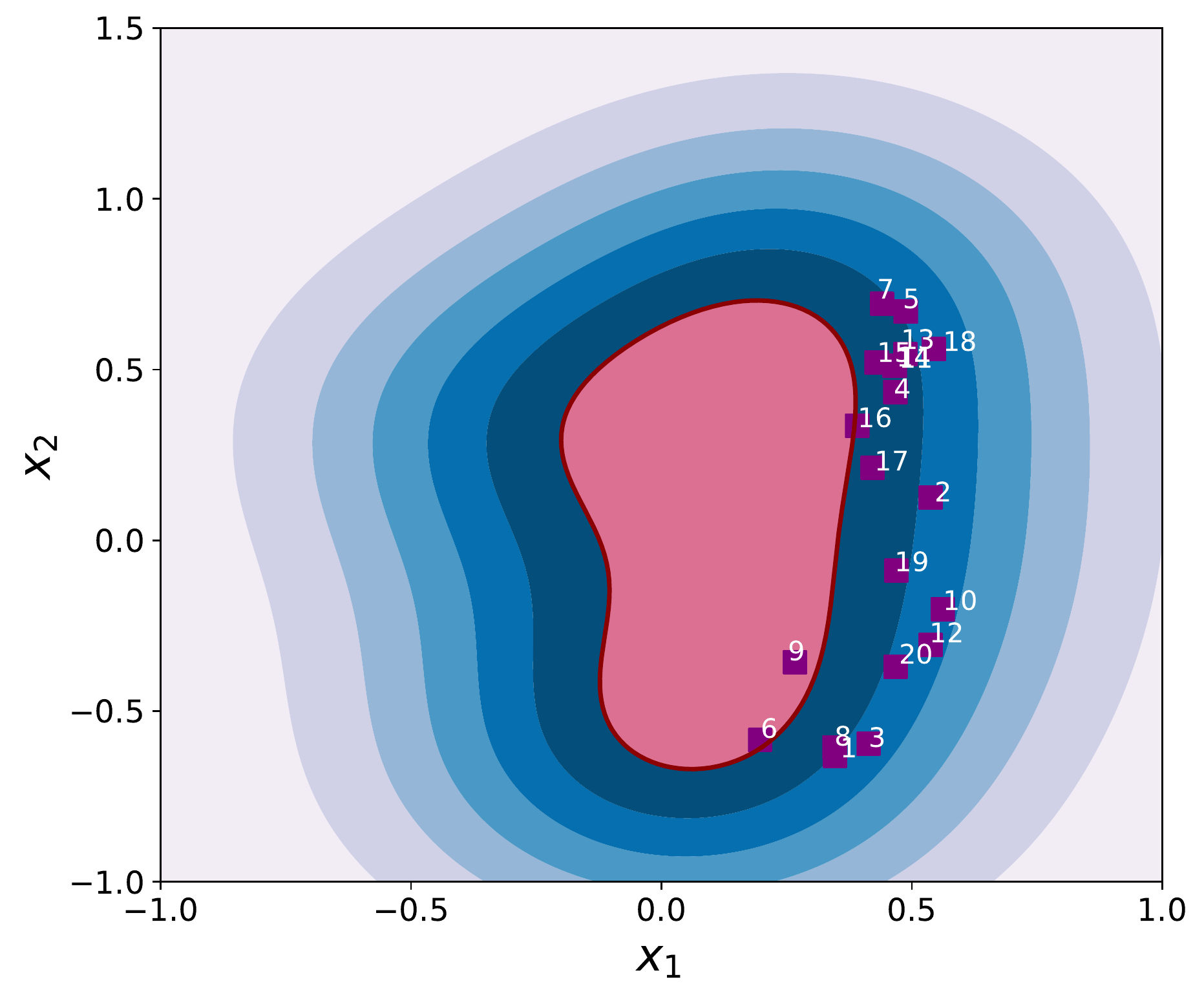}}
\subfigure[]{\includegraphics[width=0.7
 \linewidth, trim={0cm 0cm 0cm 0cm},clip]{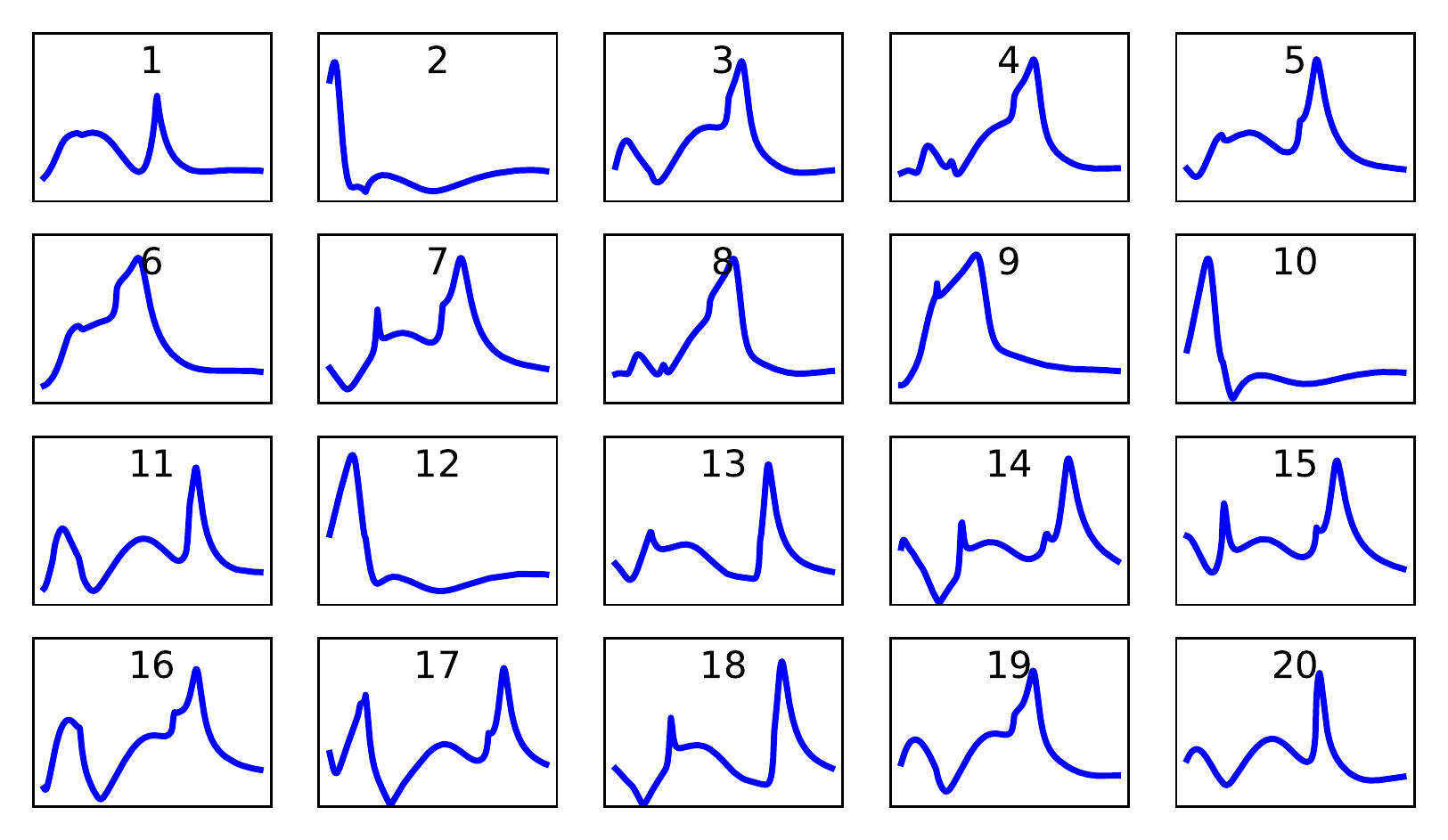}}
\caption*{Figure S7: (a) The trained one-class SVM  for the 14$\times$14 binary in Fig. 2(b). (b) Representation of the 20 random reflection responses achieved from plasmonic oligomers with 20 set of random design parameters. The corresponding number for each response is shown in the one-class SVM in (a).}
\label{fig:One-class SVM numerated}
\end{figure}

\begin{table}
    \centering
        \caption*{Table S1: Details for the trained autoencoder}
    \label{tab:AE info}
    \begin{tabular}{|c|c|}
    \hline
	 Activation Function & Tangent Hyperbolic\\
	\hline

    Training Data Division & mini-batches (batch size=200) \\
	\hline
    Optimizer & Adam\\
	\hline
    Loss Function & MSE \\
    \hline
	
\end{tabular}

\end{table}

\clearpage
\newpage

\bibliography{sample.bib}

\bibliographystyle{ieeetr}

\end{document}


\baselineskip24pt


\maketitle


\begin{sciabstract}
 Here, we present a new approach based on geometric deep learning to measure the viability of a certain optical response from a class of nanostructures. Design of nanostructures always comes with having some constraints (e.g. size, shape, and material properties). Due to these constraints, some responses are not practical for a certain class of nanostructure by any means (any set of design parameters). In case a given response is impossible to achieve using a class of nanostructure, instead of searching blindly over all possible design parameters, which is expensive in time and resource, this approach leads us to change the structure. The algorithm first reduces the dimensionality of the response space using the ground truth data generated by commercial full-wave simulators. Then, we train the platform to find the optimum convex-hull to bound the feasible responses in the latent space. This is done through an iterative process until convergence. Next, we apply one-class SVM algorithm to find the non-convex geometry of achievable responses. As a proof of concept, we apply our method to two different classes of metasurfaces (MSs)  in the visible range: i) digital MSs consisting of 7x7 and 14x14 binary plasmonic nano-cubes associated with sophisticated numerical reflectance responses, and ii) engineered MSs comprising of a square-lattice array of dielectric nano-ellipsoids associated with numerical and experimental Fano-type sharp resonances. We envision the algorithm can accelerate current design approaches and grant priceless information about what a specific class of photonics nanostructure is capable to offer.
\end{sciabstract}

% In setting up this template for *Science* papers, we've used both
% the \section* command and the \paragraph* command for topical
% divisions.  Which you use will of course depend on the type of paper
% you're writing.  Review Articles tend to have displayed headings, for
% which \section* is more appropriate; Research Articles, when they have
% formal topical divisions at all, tend to signal them with bold text
% that runs into the paragraph, for which \paragraph* is the right
% choice.  Either way, use the asterisk (*) modifier, as shown, to

\section*{Introduction}

Photonic nanostructures have been of great recent interest due to their unique capabilities to manipulate the properties of electromagnetic (EM) waves beyond what conventional bulk materials can do. Owing to their constituent nanoscale features, which can spectrally, spatially, or temporally control the optical state of EM waves with subwavelength resolution, nanophotonic devices extend all the functionalities realized by conventional bulky optical devices in much smaller footprints \cite{ding2017gradient,kamali2018review,ding2019dynamic,genevet2017recent,jahani2016all,zhan2016low,jiang2019metasurface}. Combined with the advances in nanofabrication technologies, these nanostructures have been used to demonstrate devices with enormous potential for groundbreaking technologies such as computing \cite{chizari2016analog,shen2017deep,zhu2017plasmonic,abdollahramezani2017dielectric}, imaging \cite{colburn2018varifocal,huang2013three,colburn2018metasurface}, and energy harvesting \cite{liu2011taming,ding2014ultrabroadband}, to name a few. 

Design of photonic devices in the nanoscale regime outperforming the bulky optical components has been a long-lasting challenge in some state-of-the-art applications. Accordingly, devising a comprehensive model to understand and explain the fundamental physics of light-matter interactions in these nanostructures is a substantial step toward the realization of novel nanophotonic devices. To this end, existing modeling methods can be categorized into two main groups; single- and multi-objective approaches. Single-objective approaches either rely on exhaustive design parameter sweeps using a brute-force EM solver (e.g., based on the finite element method) \cite{wu2003design} or evolve from an initial guess to a final result through evolutionary methods (e.g., genetic algorithm) \cite{bossard2014near}. While the former requires extensive computation, the latter highly depends on the initial guess and in most cases converges to a local optimum. Both of these single-objective approaches are computationally demanding and fail when the input-output relation is complex, or the number of desired features for a nanostructure grows. On the other hand, multi-objective methods \cite{jiang2019global,kiarashinejad2019deepdim} deal with formation of a model to optimize a certain class of problems. Although these methods are more computationally efficient, obtaining an optimal solution is not guaranteed.